# A singular perturbation study of the Rolie-Poly model

Yuriko Renardy and Michael Renardy

*Department of Mathematics, 460 McBryde Hall, 225 Stanger Street, Virginia Tech, Blacksburg, VA 24061-0123*

**Abstract**

We study the Rolie-Poly model for entangled polymers, using a singular perturbation analysis for the limit of large relaxation time. In this limit, it is shown that the model displays the characteristic features of thixotropic yield stress fluids, including yield stress hysteresis, delayed yielding and long term persistence of a decreased viscosity after cessation of flow. We focus on the startup and cessation of shear flow. We identify dynamic regimes of fast, slow and yielded dynamics, and show how the combination of these regimes can be used to describe the flow.

## 1. Introduction

Thixotropic yield stress fluids display characteristics that cannot be explained by "simple" yield stress models like the Bingham or Herschel-Bulkley model, such as yield stress hysteresis, delayed yielding, and long term persistence of a decreased viscosity even after cessation of flow. In the literature, this is often modeled by introducing a structure parameter, on which viscosity and yield stress depend, and which evolves on a slow time scale. In [6], an alternative approach is developed which obtains thixotropic behavior as a singular limit of viscoelasticity with a large relaxation time. The underlying model is Larson's [3] Partially Extending Strand Convection (PEC) model, which was originally proposed as a model for entangled polymers, but has more recently been modified to describe wormlike micelles [8]. This model, strictly speaking, has a yield stress only in the limit of infinite relaxation time. For large but finite relaxation time, there is a viscosity jump of several orders of magnitude; this is sometimes referred to as “apparent” yield stress behavior. The analysis of [6] shows how the model predicts yield stress hysteresis, delayed yielding, and thixotropy.

The goal of this paper is a similar analysis for another model for entangled polymers, known as the Rolie-Poly model. This model has recently gained popularity because it can predict transient shear banding, even in situations where the steady shear stress vs. shear rate behavior is monotone [1, 2]. We link this observation to delayed yielding below. The Rolie-Poly model is described in [4]. It uses two relaxation times: $\tau_d$, which is associated with reptation, and the Rouse time $\tau_R$. The stress tensor is given by $\mathbf{T} = G\mathbf{C}$, where $\mathbf{C}$ is a conformation tensor satisfying the constitutive equation

$$\dot{\mathbf{C}} - (\nabla\mathbf{v})\mathbf{C} - \mathbf{C}(\nabla\mathbf{v})^T = -\frac{1}{\tau_d}(\mathbf{C}-\mathbf{I}) - \frac{2(1-\sqrt{3/\operatorname{tr}\mathbf{C}})}{\tau_R}(\mathbf{C}+\beta(\frac{\operatorname{tr}\mathbf{C}}{3})^\delta(\mathbf{C}-\mathbf{I})). \tag{1}$$

Here $\dot{\mathbf{C}}$ denotes the material time derivative of $\mathbf{C}$. (In this paper, we address homogeneous flows only, and thus the material time derivative reduces to the ordinary time derivative.) The convective constraint release (CCR) coefficient $\beta$ and the exponent $\delta$ are dimensionless model parameters. The ratio $\tau_R/\tau_d$ is small, and this will provide the small parameter, denoted by $\epsilon$, for our analysis below. We nondimensionalize time in units of $\tau_R$, and stress in units of $G$. The dimensionless quantities (with asterisks) are $\mathbf{T}^* = \mathbf{T}/G$, $t^* = t/\tau_R$, $\nabla^*\mathbf{v}^* = \tau_R\nabla\mathbf{v}$. Hereafter, we drop the asterisks. Thus the nondimensional form of the constitutive equation becomes

$$\dot{\mathbf{C}} - (\nabla\mathbf{v})\mathbf{C} - \mathbf{C}(\nabla\mathbf{v})^T = -\epsilon(\mathbf{C}-\mathbf{I}) - 2(1-\sqrt{3/\operatorname{tr}\mathbf{C}})(\mathbf{C}+\beta(\frac{\operatorname{tr}\mathbf{C}}{3})^\delta(\mathbf{C}-\mathbf{I})). \tag{2}$$

Before we analyze the full Rolie-Poly model, we investigate a simpler “non-stretching” version of the model, where a limit $\tau_R \to 0$ is taken in such a way that $\operatorname{tr}\mathbf{C} \to 3$ at the same rate at which $\tau_R \to 0$. This model is obtained by taking the trace in (1):

$$\begin{aligned} \frac{d}{dt}(\operatorname{tr}\mathbf{C}) &= 2\operatorname{tr}((\nabla\mathbf{v})\mathbf{C}) - \frac{1}{\tau_d}(\operatorname{tr}\mathbf{C}-3) \\ &\quad -\frac{2}{\tau_R}(1-\sqrt{\frac{3}{\operatorname{tr}\mathbf{C}}})(\operatorname{tr}\mathbf{C}+\beta(\frac{\operatorname{tr}\mathbf{C}}{3})^\delta(\operatorname{tr}\mathbf{C}-3)), \end{aligned} \tag{3}$$

setting $\operatorname{tr}\mathbf{C} = 3 + \tau_R\Delta$, and ignoring all terms of order $\tau_R$. This yields

$$0 = 2\operatorname{tr}((\nabla\mathbf{v})\mathbf{C}) - \Delta. \tag{4}$$

Using this in (1), and again ignoring terms which vanish as $\tau_R \to 0$, we obtain the non-stretching form of the Rolie-Poly model,

$$\dot{\mathbf{C}} - (\nabla\mathbf{v})\mathbf{C} - \mathbf{C}(\nabla\mathbf{v})^T = -\frac{2}{3}\operatorname{tr}((\nabla\mathbf{v})\mathbf{C})(\mathbf{C} + \beta(\mathbf{C} - \mathbf{I})) - \frac{1}{\tau_d}(\mathbf{C} - \mathbf{I}). \tag{5}$$

A Newtonian term is included in the constitutive equation, so that the extra stress tensor is

$$\mathbf{T} = G\mathbf{C} + \eta(\nabla\mathbf{v} + (\nabla\mathbf{v})^T). \tag{6}$$

For nondimensionalization, we scale stress with $G$, time with the retardation time $\eta/G$, and define the small parameter by $\epsilon = \eta/(G\tau_d)$. Thus, in dimensionless form, the stress is given by $\mathbf{T} = \mathbf{C} + \nabla\mathbf{v} + (\nabla\mathbf{v})^T$, and the constitutive equation is

$$\dot{\mathbf{C}} - (\nabla\mathbf{v})\mathbf{C} - \mathbf{C}(\nabla\mathbf{v})^T = -\frac{2}{3}\operatorname{tr}((\nabla\mathbf{v})\mathbf{C})(\mathbf{C} + \beta(\mathbf{C} - \mathbf{I})) - \epsilon(\mathbf{C} - \mathbf{I}). \tag{7}$$

Section 2 concerns the behavior of the nonstretching model in the limit of small $\epsilon$. The startup of shear flow under a prescribed shear stress is discussed. Distinct regimes of fast, slow and yielded dynamics are found, as in the PEC model [6]. If the imposed shear stress is small enough, the initial fast dynamics transitions to slow dynamics and reaches an unyielded equilibrium. For large imposed stress, fast dynamics transitions to yielded dynamics, and a yielded equilibrium is reached. There is an intermediate range, however, where delayed yielding occurs: there is a transition from fast to slow dynamics, but no equilibrium is reached in slow dynamics. Instead, there is ultimately another transition to fast dynamics, followed by yielded dynamics. This delayed yielding occurs even if the constitutive curve for steady shear is monotone. The steady flow that is ultimately reached is yielded, but yielding occurs after a long time. If the shear stress is slightly inhomogeneous, as is inevitable in a real device, this inhomogeneity is magnified in the yield time, and transient shear bands form [1, 2].

We also study cessation of shear flow, where the initial state is an established steady shear flow, and then the imposed shear stress is reduced to zero. In this case, we find, as in [6], that the motion comes to a stop quickly, but $\mathbf{C}$ relaxes back to its equilibrium value of $\mathbf{I}$ over a slow time scale. The result of this is thixotropic behavior.

Section 3 provides a similar analysis for the stretching Rolie-Poly model. We focus on the case $\delta = 0$, since only this case seems to be accessible to a fully analytical treatment. The results of the analysis show analogies to the nonstretching case. Section 4 is a discussion of numerical results for cases with nonzero $\delta$. The behavior is interpreted in the light of results from the analysis for $\delta = 0$.

## 2. The non-stretching Rolie-Poly model

The non-stretching Rolie-Poly model is a kind of twin to the PEC model studied in [6]; in the PEC model, we have $\beta = 0$, and instead there is a linear combination of $\mathbf{C}$ and $\mathbf{I}$ in the $\epsilon$-term. The non-stretching Rolie-Poly with $\beta = 0$ is equivalent to the PEC model with (in the notation of [6]) $\alpha = 0$. In shear flow, we have

$$\mathbf{C} = \begin{pmatrix} C_{11} & C_{12} & 0 \\ C_{12} & C_{22} & 0 \\ 0 & 0 & C_{22} \end{pmatrix}, \tag{8}$$

and, with $\kappa$ denoting the shear rate, we get

$$\begin{aligned} \dot{C}_{11} &= 2\kappa C_{12} - \frac{2}{3}\kappa C_{12}((1+\beta)C_{11} - \beta) - \epsilon(C_{11} - 1), \\ \dot{C}_{12} &= \kappa C_{22} - \frac{2}{3}(1+\beta)\kappa C_{12}^2 - \epsilon C_{12}, \\ \dot{C}_{22} &= -\frac{2}{3}\kappa C_{12}((1+\beta)C_{22} - \beta) - \epsilon(C_{22} - 1). \end{aligned} \tag{9}$$

If the total shear stress $\tau$ is prescribed, then $\kappa = \tau - C_{12}$ (recall our nondimensionalization, which sets the Newtonian part of the viscosity equal to 1).

We can combine the equations above to yield

$$\frac{d}{dt}(C_{11} + 2C_{22} - 3) = -(\frac{2}{3}\kappa C_{12}(1+\beta) + \epsilon)(C_{11} + 2C_{22} - 3). \tag{10}$$

Therefore $C_{11} + 2C_{22} = 3$ if this is the case initially; this is why the model is called "non-stretching". Henceforth, we focus on the evolution of $C_{12}$ and $C_{22}$.

*2.1. Steady shear flow*

For steady shear flow, we have

$$\begin{aligned} \kappa C_{22} - \frac{2}{3}(1+\beta)\kappa C_{12}^2 - \epsilon C_{12} &= 0, \\ -\frac{2}{3}\kappa C_{12}((1+\beta)C_{22} - \beta) - \epsilon(C_{22} - 1) &= 0. \end{aligned} \tag{11}$$

We solve the first equation for $\kappa$, which results in

$$\kappa = \frac{3C_{12}\epsilon}{3C_{22} - 2(1+\beta)C_{12}^2}. \tag{12}$$

We use this in the second equation and solve for $C_{12}$. This yields

$$C_{12} = \pm\sqrt{\frac{3}{2}}\sqrt{C_{22} - C_{22}^2}. \tag{13}$$

The expression for $\kappa$ now becomes

$$\kappa = \pm\frac{\sqrt{\frac{3}{2}}\sqrt{C_{22}(1 - C_{22})}\epsilon}{C_{22}((1+\beta)C_{22} - \beta)}. \tag{14}$$

This leads to the following conclusions:

1. Physically relevant steady flows exist for $C_{22}$ in the range $1 \geq C_{22} > \beta/(1+\beta)$. As $C_{22}$ varies from 1 to $\beta/(1+\beta)$, $\kappa$ varies from 0 to $\pm\infty$ and $C_{12}$ varies from 0 to $\pm\sqrt{3\beta/2}/(1+\beta)$.
2. In the range $0 \leq C_{22} < \beta/(1+\beta)$, there exist solutions for which $\kappa$ and $C_{12}$ have opposite signs. These are an artifact of the model and will not be discussed further.
3. The elastic shear stress $C_{12}$ goes through a maximum when $C_{22} = 1/2$, while the shear rate $\kappa$ is a monotone function of $C_{22}$. Hence the model has a nonmonotone shear stress versus shear rate curve if $\beta < 1$, i.e. $1/2 > \beta/(1+\beta)$; it is monotone for $\beta \geq 1$. For $\beta < 1$, and sufficiently small $\epsilon$, the total shear stress $\tau = C_{12} + \kappa$ has a maximum at a shear rate of order $\epsilon$ and a minimum at a shear rate of order $\sqrt{\epsilon}$.

Figure 1 shows the steady flow curve for $\epsilon = 0.001$ and for the two values $\beta = 0.1$ and $\beta = 1.5$.

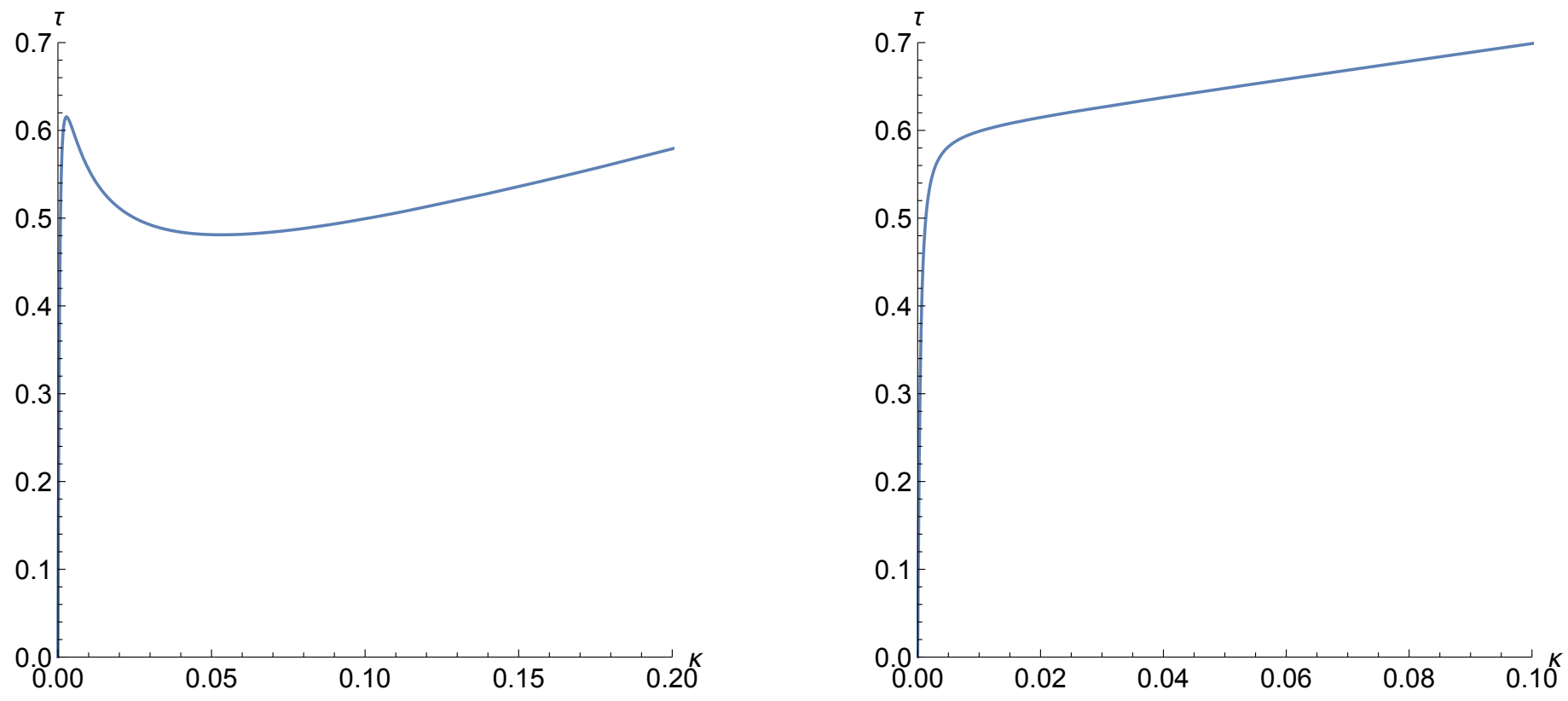

Figure 1: Steady flow curves for $\epsilon = 0.001$ and $\beta = 0.1$, $\beta = 1.5$.

*2.2. Fast dynamics*

We consider startup of shear flow from rest, under an imposed constant shear stress $\tau$. That is, we have $\kappa = \tau - C_{12}$, and the initial condition is $C_{12} = 0$, $C_{22} = 1$. If $\epsilon$ is small, we can, at least initially, approximate the dynamics by setting $\epsilon = 0$. We refer to this regime as fast dynamics.

We introduce the strain $\gamma$, defined as the integral $\int_0^t \kappa(s)\,ds$. The equations for fast dynamics then become

$$\begin{aligned} \frac{dC_{12}}{d\gamma} &= C_{22} - \frac{2}{3}(1+\beta)C_{12}^2, \\ \frac{dC_{22}}{d\gamma} &= -\frac{2}{3}C_{12}((1+\beta)C_{22} - \beta). \end{aligned} \tag{15}$$

We use the notation $q = C_{12}/C_{22}$. The two equations of (15) can be combined to yield

$$\frac{dq}{d\gamma} = 1 - \frac{2}{3}\beta q^2. \tag{16}$$

With the initial condition $q(0) = 0$, this leads to

$$q(\gamma) = \sqrt{\frac{3}{2\beta}}\tanh(\sqrt{\frac{2\beta}{3}}\gamma). \tag{17}$$

We can now replace $C_{12}$ in the second equation of (15) by $qC_{22}$ and solve for $C_{22}$. In this fashion we obtain

$$C_{22}(\gamma) = \frac{\beta}{1+\beta-\operatorname{sech}(\sqrt{\frac{2\beta}{3}}\gamma)},$$

$$C_{12}(\gamma) \;=\; \frac{\sqrt{\frac{3\beta}{2}}\sinh(\sqrt{\frac{2\beta}{3}}\gamma)}{(1+\beta)\cosh(\sqrt{\frac{2\beta}{3}}\gamma)-1}. \tag{18}$$

For $\gamma \to \infty$, $C_{22}(\gamma)$ approaches $\beta/(1+\beta)$, and $C_{12}(\gamma)$ approaches $\sqrt{3\beta/2}/(1+\beta)$ from above. Before that, $C_{12}(\gamma)$ reaches a maximum at

$$\gamma = \sqrt{\frac{3}{2\beta}}\ln(1+\beta-\sqrt{2\beta+\beta^2}). \tag{19}$$

The value of the maximum is $\sqrt{3/(2(2+\beta))}$.

We note, however, that when $C_{12}$ reaches $\tau$, then $\kappa$ becomes zero. When $\kappa$ becomes small of order $\epsilon$, the assumption of neglecting the order $\epsilon$ terms in the equations loses its validity and fast dynamics no longer applies. In this case, we have a transition to slow dynamics, described in the next section. We thus find the following two cases:

1. If $\tau > \sqrt{3/(2(2+\beta))}$, then $\kappa$ remains positive until $\gamma \to \infty$. The solution reaches yielded behavior where $C_{22} = \beta/(1+\beta)$ and $C_{12} = \sqrt{3\beta/2}/(1+\beta)$.
2. If $\tau < \sqrt{3/(2(2+\beta))}$, then $C_{12}$ reaches $\tau$, and a transition to slow dynamics occurs. The value of $C_{22}$ at which this occurs is

$$C_{22} = \frac{3(1+\beta)+\sqrt{3}\sqrt{3-2(2+\beta)\tau^2}}{3(2+\beta)}. \tag{20}$$

*2.3. Slow dynamics*

In the slow dynamics regime, $\kappa$ is small on the same order as $\epsilon$, and the time evolution is on a slow time scale. Accordingly, we set $\kappa = \tilde{\kappa}\epsilon$, $C_{12} = \tau - \tilde{\kappa}\epsilon$, and $t = \tilde{t}/\epsilon$. At leading order, we find

$$\begin{aligned} 0 &= \tilde{\kappa}C_{22} - \frac{2}{3}(1+\beta)\tilde{\kappa}\tau^2 - \tau, \\ \frac{dC_{22}}{d\tilde{t}} &= -\frac{2}{3}\tilde{\kappa}\tau((1+\beta)C_{22}-\beta) - (C_{22}-1). \end{aligned} \tag{21}$$

We can solve the first equation for $\tilde{\kappa}$ and obtain a first order ODE for $C_{22}$. This ODE takes the form

$$\frac{dC_{22}}{d\tilde{t}} = \frac{3C_{22}-3C_{22}^2-2\tau^2}{3C_{22}-2(1+\beta)\tau^2}. \tag{22}$$

If $\tau^2 \leq 3/8$, there are two equilibrium points given by the zeros of the numerator:

$$C_{22} = \frac{1}{2} \pm \frac{1}{2\sqrt{3}}\sqrt{3 - 8\tau^2}. \tag{23}$$

The denominator of (22) vanishes when

$$C_{22} = 2(1+\beta)\tau^2/3 =: C_{22}^s, \tag{24}$$

This is a point where the slow curve (given by $C_{12} = \tau$) is tangent to a fast curve given by (15): If we set $C_{12} = \tau$, the right hand side of the first equation in (15) vanishes precisely if $C_{22} = 2(1+\beta)\tau^2/3$.

We denote the value given by (20) by $C_{22}^i$, the larger of the two equilibrium values given by (23) by $C_{22}^{eq}$ and the singular value given by (24) by $C_{22}^s$. Below, we establish the following inequalities between these values:

1. If either $0 \leq \beta < 1$ and $0 < \tau < \sqrt{3/8}$ or $\beta > 1$ and $0 < \tau < \sqrt{3\beta/2}/(1+\beta)$, then $C_{22}^i > C_{22}^{eq} > C_{22}^s$.
2. If $\beta < 1$ and $\sqrt{3/8} < \tau < \sqrt{3/(2(2+\beta)}$, then $C_{22}^i > C_{22}^s$, and $C_{22}^{eq}$ does not exist.
3. If $\beta > 1$, and $\sqrt{3\beta/2}/(1+\beta) < \tau < \sqrt{3/2(2+\beta)}$, then $C_{22}^i > C_{22}^s$, and $C_{22}^s > C_{22}^{eq}$ if $C_{22}^{eq}$ exists.

First of all, we check that, for small $\tau$, we have

$$C_{22}^{eq} \sim 1 - \frac{2}{3}\tau^2, \quad C_{22}^i \sim 1 - \frac{1}{3}\tau^2, \quad C_{22}^s \sim \frac{2}{3}(1+\beta)\tau^2, \tag{25}$$

so clearly $C_{22}^i > C_{22}^{eq} > C_{22}^s$. We next investigate when $C_{22}^{eq} - C_{22}^s$ changes sign. For the equation $C_{22}^{eq} = C_{22}^s$, Mathematica yields the solution $\tau = \sqrt{3\beta/2}/(1+\beta)$. However, this solution pays no attention to the sign of square roots, and a further check reveals that $C_{22}^{eq} - C_{22}^s$ changes sign at this point only if $\beta > 1$. The equation $C_{22}^i = C_{22}^{eq}$ also yields the formal roots $\tau = \sqrt{3\beta/2}/(1+\beta)$. For $\beta < 1$, we find that, at this point, we have actually

$$C_{22}^i = \frac{2+2\beta+\beta^2}{(1+\beta)(2+\beta)} > C_{22}^{eq} = \frac{1}{1+\beta} > C_{22}^s = \frac{\beta}{1+\beta}. \tag{26}$$

On the other hand, for $\beta > 1$, we have

$$C_{22}^i = \frac{2+2\beta+\beta^2}{(1+\beta)(2+\beta)} > C_{22}^s = C_{22}^{eq} = \frac{\beta}{1+\beta}. \tag{27}$$

The other formal root of the equation $C^i_{22} = C^s_{22}$ is $\tau = \sqrt{3/2(2+\beta)}$. At this point, we have $C^i_{22} = C^s_{22} = (1+\beta)/(2+\beta)$, so there is a change of sign. For this value of $\tau$, $C^{eq}_{22}$ does not exist if $\beta < 2$, and we have

$$C^{eq}_{22} = \frac{1}{2}(1 + \sqrt{\frac{\beta-2}{\beta+2}}) < C^s_{22} \tag{28}$$

if $\beta \geq 2$.

Now we consider solutions starting from equilbrium along a fast curve and entering the slow dynamics under alternative 2 outlined at the end of the previous subsection. They will enter the slow manifold at $C_{22} = C^i_{22}$. According to the lemma, we have the following two alternatives.

1. If $0 \leq \beta < 1$ and $0 < \tau < \sqrt{3/8}$ or $\beta > 1$ and $0 < \tau < \sqrt{3\beta/2}/(1+\beta)$, the equilibrium point $C^{eq}_{22}$ is ultimately reached via slow dynamics. This is unyielded behavior.
2. If $\beta < 1$ and $\sqrt{3/8} < \tau < \sqrt{3/(2(2+\beta))}$, or $\beta > 1$, and $\sqrt{3\beta/2}/(1+\beta) < \tau < \sqrt{3/(2(2+\beta))}$, the singular point is reached along the slow manifold. At this point, the dynamics transitions to another fast curve, along which yielded equilibrium is reached. This is the case of delayed yielding. The delay time is essentially equal to the residence time along the slow manifold, i.e. it is of order $1/\epsilon$.

*2.4. Startup of shear flow from rest*

In summary, the results of the preceding subsections yield the following behaviors in startup of shear flow from rest.

*2.4.1. Case $\beta < 1$*

1. If $0 < \tau < \sqrt{3/8}$, the flow reaches unyielded equilbrium.
2. If $\sqrt{3/8} < \tau < \sqrt{3/(2(2+\beta))}$, delayed yielding occurs.
3. If $\tau > \sqrt{3/(2(2+\beta))}$, immediate yielding occurs.

*2.4.2. Case $\beta > 1$*

1. If $0 < \tau < \sqrt{3\beta/2}/(1+\beta)$, the flow reaches unyielded equilbrium.
2. If $\sqrt{3\beta/2}/(1+\beta) < \tau < \sqrt{3/(2(2+\beta))}$, delayed yielding occurs.
3. If $\tau > \sqrt{3/(2(2+\beta))}$, immediate yielding occurs.

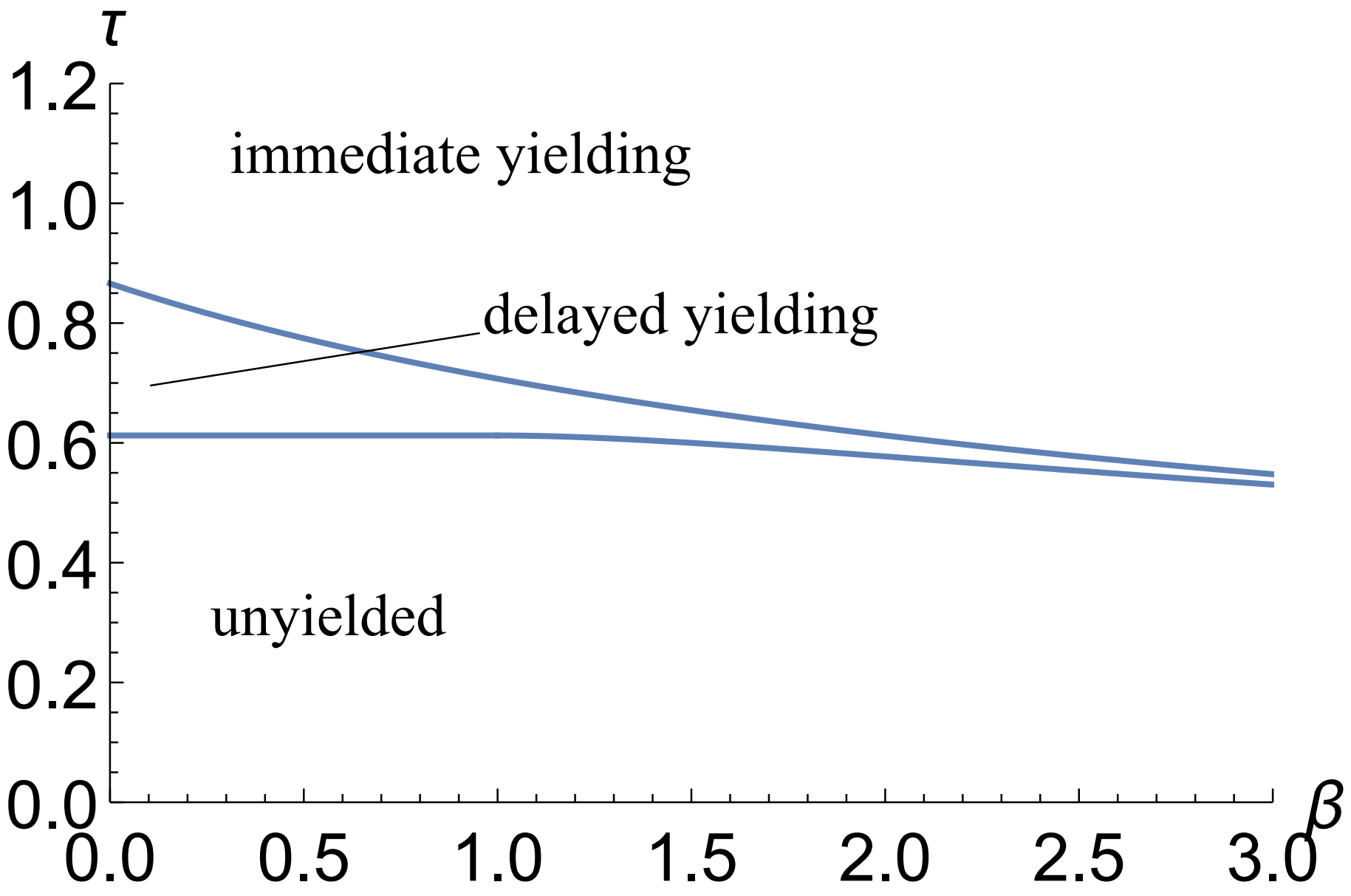

Figure 2: Regions of unyielded behavior, delayed yielding and immediate yielding in startup of shear flow for the nonstretching Rolie-Poly model.

Figure 2 shows the regions of unyielded behavior, delayed yielding and immediate yielding in the $(\beta, \tau)$-plane.

Figure 3 shows the evolution of the shear rate as a function of time for $\beta = 0$ and various values of $\tau$. As $\tau$ is increased from $0.6 < \sqrt{3/8}$ to $0.9 > \sqrt{3/4}$, we have a transition from unyielded behavior to immediate yielding, with delayed yielding at values in between. The shear rate for Figure 3 (a) settles to a value of approximately $0.001{=}O(\epsilon)$ for $t > 10$. Part (d) might give the impression that yielding is still "delayed." However, we have rerun the simulation of Figure 3 (d) with $\epsilon = 0.0001$, and the plot looks virtually identical, so there is no delay of order $1/\epsilon$.

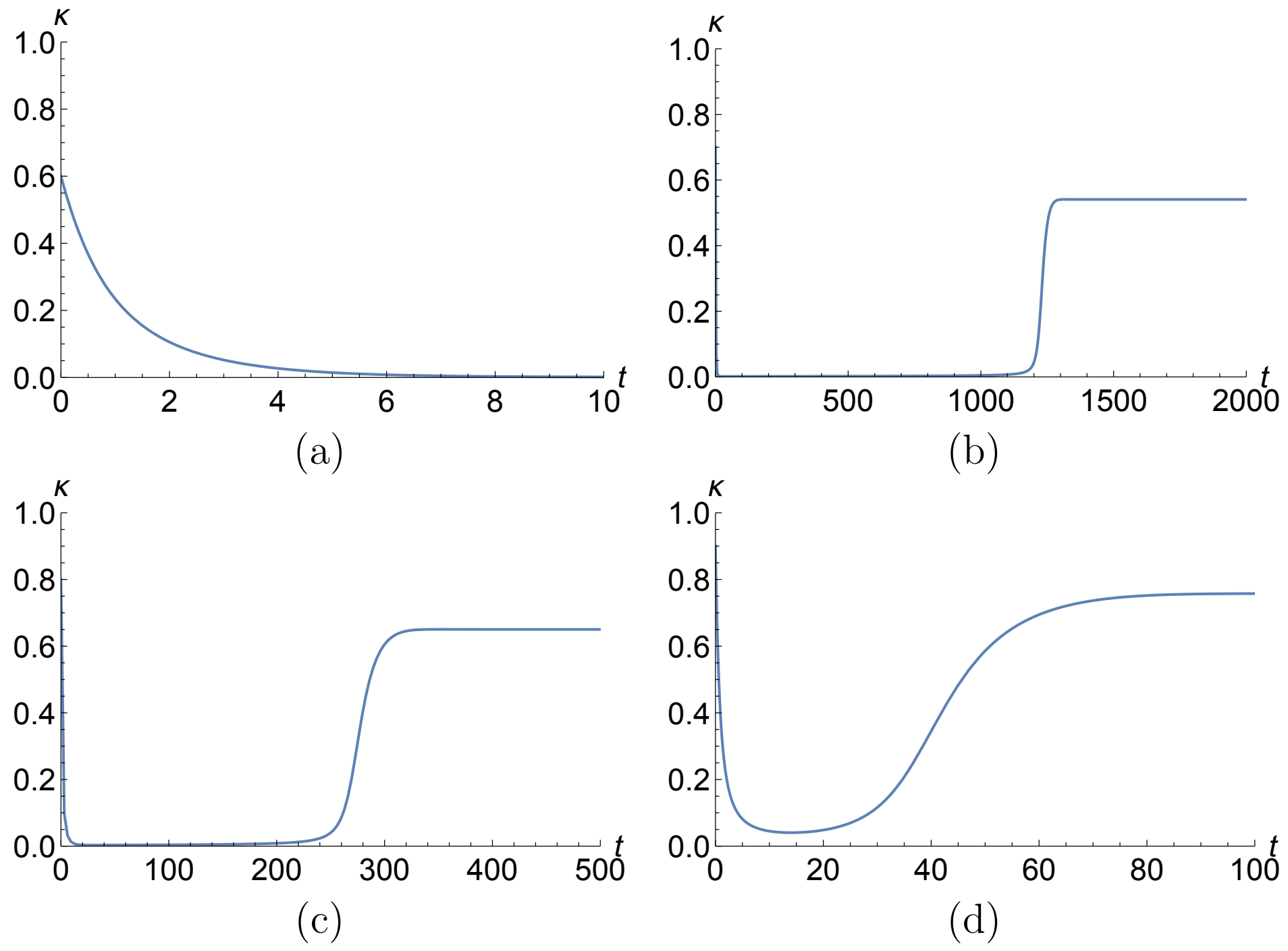

Figure 3: Evolution of shear rate for startup of shear flow for the nonstretching Rolie-Poly model, $\epsilon = 0.001$, $\beta = 0$. (a) $\tau = 0.6$, (b) $\tau = 0.7$, (c) $\tau = 0.8$, (d) $\tau = 0.9$.

For $\beta = 0.8$, the analysis predicts delayed yielding between $\tau = \sqrt{3/8} = 0.612$ and $\tau = 0.732$. This agrees well with the simulations for $\epsilon = 0.001$ shown in Figure 4.

Finally, we show a case where $\beta > 1$. For $\beta = 1.5$, the analysis predicts unyielded behavior if $\tau < 0.6$, delayed yielding if $0.6 < \tau < 0.655$, and immediate yielding for $\tau > 0.655$. Again the plots shown in Figure 5 are in good agreement.

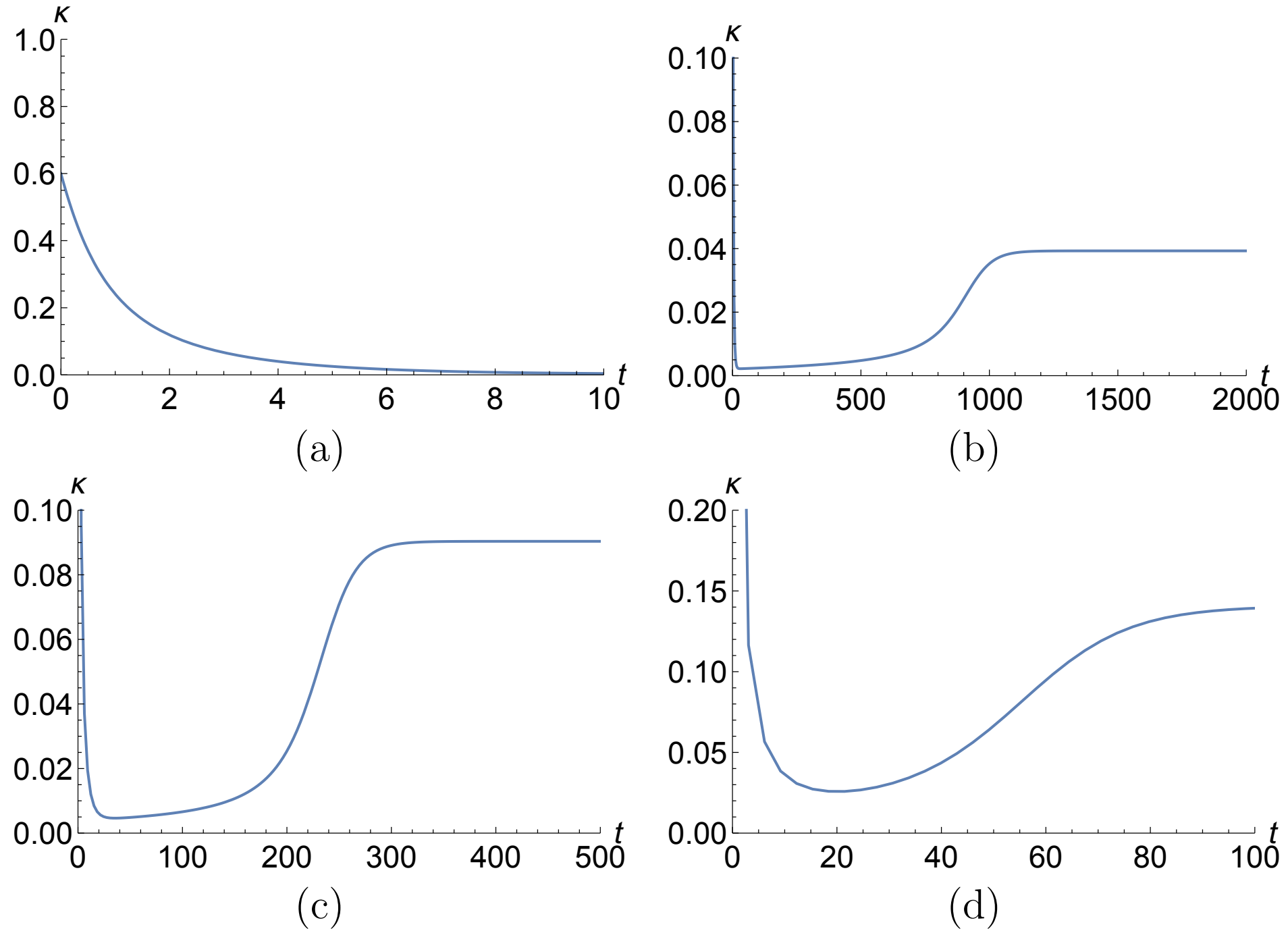

Figure 4: Evolution of shear rate for startup of shear flow for the nonstretching Rolie-Poly model, $\epsilon = 0.001$, $\beta = 0.8$. (a) $\tau = 0.6$, (b) $\tau = 0.65$, (c) $\tau = 0.7$, (d) $\tau = 0.75$.

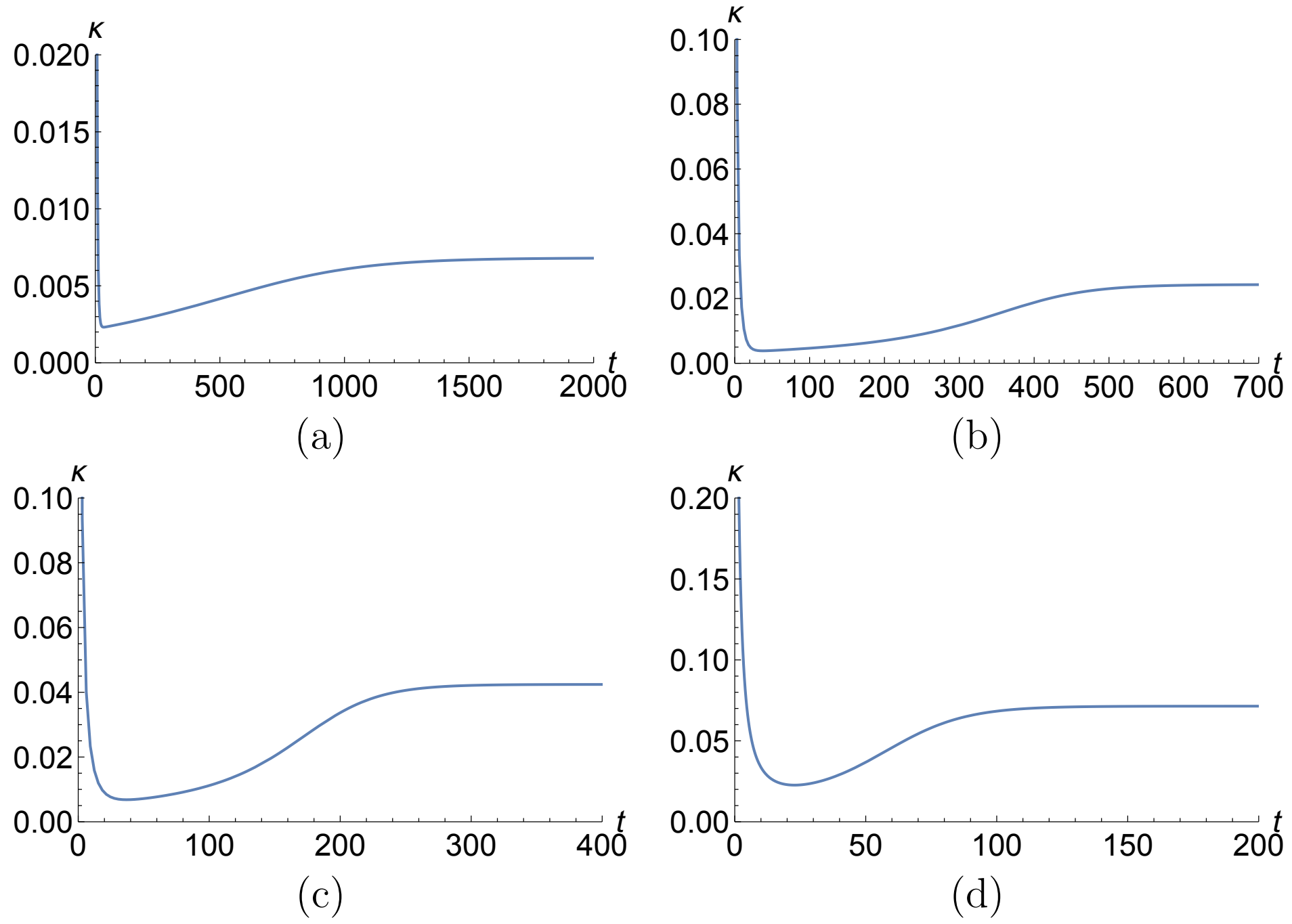

Figure 5: Evolution of shear rate for startup of shear flow for the nonstretching Rolie-Poly model, $\epsilon = 0.001$, $\beta = 1.5$. (a) $\tau = 0.59$, (b) $\tau = 0.62$, (c) $\tau = 0.64$, (d) $\tau = 0.67$.

*2.5. Yielded dynamics and unyielding*

The fast dynamics (15) has the fixed point $C_{22} = C_{22}^0 = \beta/(1+\beta)$, $C_{12} = C_{12}^0 = \sqrt{3\beta/2}/(1+\beta)$, which represents yielded steady flow. To predict details of the evolution after yielding, in particular the unyielding process when stress is reduced, a more detailed analysis of the vicinity of this fixed point is necessary. For this purpose, we expand to the next order in $\epsilon$. We set $C_{12} = C_{12}^0 + \epsilon c_{12}$, $C_{22} = C_{22}^0 + \epsilon c_{22}$, insert this ansatz into the full equations, and linearize with respect to $\epsilon$. For $\beta \neq 0$, this results in the system

$$\begin{aligned}(1+\beta)\dot{c}_{12} &= 2(\beta - \frac{1}{3}\sqrt{6\beta}\tau(1+\beta))c_{12} + ((1+\beta)\tau - \frac{1}{2}\sqrt{6\beta})c_{22} - \frac{1}{2}\sqrt{6\beta},\\ (1+\beta)\dot{c}_{22} &= (\beta - \frac{1}{3}\sqrt{6\beta}\tau(1+\beta))c_{22} + 1. \qquad (29)\end{aligned}$$

We refer to these equations as "yielded dynamics." For the specific case of the nonstretching Rolie-Poly model, the yielded dynamics is just simply the linearization at the yielded steady flow. This is different, however, for the full Rolie-Poly model discussed in the next section. The equations (29) have the fixed point

$$\begin{aligned}c_{12} &= \frac{9(1-\beta)}{2\sqrt{6}\beta(-3\sqrt{\beta}+\sqrt{6}(1+\beta)\tau)},\\ c_{22} &= \frac{3}{-3\beta+\sqrt{6\beta}(1+\beta)\tau}. \qquad (30)\end{aligned}$$

This fixed point represents yielded steady flow up to terms of order $\epsilon$. The eigenvalues of the matrix representing the linear system (29) are $\beta/(1+\beta) - \frac{1}{3}\sqrt{6\beta}\tau$ and $2(\beta/(1+\beta) - \frac{1}{3}\sqrt{6\beta}\tau)$. Therefore, the fixed point exists and is stable as long as $\tau > \sqrt{3\beta/2}/(1+\beta)$. We note that, regardless of $\tau$, the fixed point is on the line

$$c_{12} = \frac{3(1-\beta)}{2\sqrt{6\beta}}c_{22}. \qquad (31)$$

Now we consider what happens if we start from an established yielded steady flow and suddenly drop the imposed shear stress $\tau$. If the new value is still higher than $\sqrt{3\beta/2}/(1+\beta)$, then the solution of (29) will simply approach the new fixed point. If, on the other hand, the imposed stress is lowered to a value less than $\sqrt{3\beta/2}/(1+\beta)$, then the fixed point is unstable, and the solution of (29) will go to infinity.

For the further analysis, we need to know in which direction of the $(c_{12}, c_{22})$-plane it goes to infinity. Let us assume that for $t < 0$, we have steady yielded flow, and the fixed point for (29) is given by $(c_{12}^a, c_{22}^a)$. For $t > 0$, $\tau$ is set to a lower value, such that the new fixed point $(c_{12}^b, c_{22}^b)$ is unstable. Let $d_{12} = c_{12} - c_{12}^b$, $d_{22} = c_{22} - c_{22}^b$. Then the evolution for $t > 0$ is given by

$$
\begin{aligned}
(1+\beta)\dot{d}_{12} &= 2(\beta - \frac{1}{3}\sqrt{6\beta}\tau(1+\beta))d_{12} + ((1+\beta)\tau - \frac{1}{2}\sqrt{6\beta})d_{22}, \\
(1+\beta)\dot{d}_{22} &= (\beta - \frac{1}{3}\sqrt{6\beta}\tau(1+\beta))d_{22}, \qquad (32)
\end{aligned}
$$

and the initial condition is

$$
d_{12}(0) = c_{12}^a - c_{12}^b, \quad d_{22}(0) = c_{22}^a - c_{22}^b. \qquad (33)
$$

We have $c_{22}^a > 0$, $c_{22}^b < 0$, so $d_{22}(0) > 0$. Moreover,

$$
d_{12}(0) = \frac{3(1-\beta)}{2\sqrt{6\beta}} d_{22}(0). \qquad (34)
$$

In the phase plane for (32), the origin is an unstable node; the fast direction is given by the $d_{12}$ axis, and the slow direction can be shown to be given by the line $d_{12} = \sqrt{6}/(2\sqrt{\beta})d_{22}$. The initial point given by (34) is always to the left of the slow manifold in the $(d_{12}, d_{22})$-plane. It follows that for $t \to \infty$, the solution of (32) will be such that $d_{12}$ is negative and $|d_{12}| >> |d_{22}|$.

The system (32) agrees with the linearization of the fast dynamics at the yielded fixed point $C_{12} = \sqrt{3\beta/2}/(1+\beta)$, $C_{22} = \beta/(1+\beta)$. The solution of (29) which we just discussed naturally matches to the line $C_{22} = \beta/(1+\beta)$ for the fast dynamics. Upon unyielding, the solution follows this line until either $C_{12} = \tau$ is reached or, if $\tau$ is sufficiently negative, $C_{12}$ approaches $-\sqrt{3\beta/2}/(1+\beta)$, i.e. there is a yielded flow in the opposite direction.

At the point where $C_{12} = \tau$ is reached, $C_{22}$ is still close to $\beta/(1+\beta)$. It will then slowly relax to its equilibrium value following the slow manifold, i.e. over a time scale of order $1/\epsilon$.

### *2.6. Thixotropy*

Let us consider in more detail the unyielding along the slow manifold. Specifically, we assume the applied stress has been reduced to zero, and $C_{22}$ is slowly increasing from $\beta/(1+\beta)$ to its equilibrium value of 1. Now at some point we reimpose a small stress $\tau$. On the slow manifold, for small $\tau$, we have approximately $\tilde{\kappa} = \tau/C_{22}$, i.e. $\kappa = \epsilon\tau/C_{22}$. This yields an apparent viscosity of $\tau/\kappa = C_{22}/\epsilon$. As $C_{22}$ increases from $\beta/(1+\beta)$ back to 1, this apparent viscosity is slowly increasing. We note that the thixotropy is much less pronounced than it is for the PEC model; unless $\beta$ is very small, the change in apparent viscosity during unyielding (from $\beta/((1+\beta)\epsilon)$ to $1/\epsilon$) is only by a factor of order 1 rather than several orders of magnitude as in [5].

If we reimpose a larger shear stress, the fluid might yield again. In Figure 6 we show phase plane pictures for the fast dynamics given by (15). If the fluid is unyielding along the slow manifold, and we then reimpose a shear stress, we start from a point close to the $C_{22}$ axis. The smaller the initial value of $C_{22}$ is, the smaller the maximum value of $C_{12}$ that is reached. Indeed, we can show that if the initial condition for fast dynamics is $C_{12} = 0$, $C_{22} = p$, then the maximum value of $C_{12}$ that is reached is

$$\sqrt{\frac{3}{2(-\beta + 2(1+\beta)p)}}p. \tag{35}$$

This value represents the threshold for immediate yielding, see the discussion of fast dynamics above. In the range $p \in [\beta/(1+\beta), 1]$, this is an increasing function of $p$. If there is a transition to slow dynamics, the value of $C_{22}$ at which this happens also decreases with decreasing initial value of $C_{22}$. That is, if delayed yielding occurs, the delay time will decrease as the initial value of $C_{22}$ decreases.

During the unyielding process, the fluid therefore "effectively" has a viscosity and yield stress that increase gradually with aging.

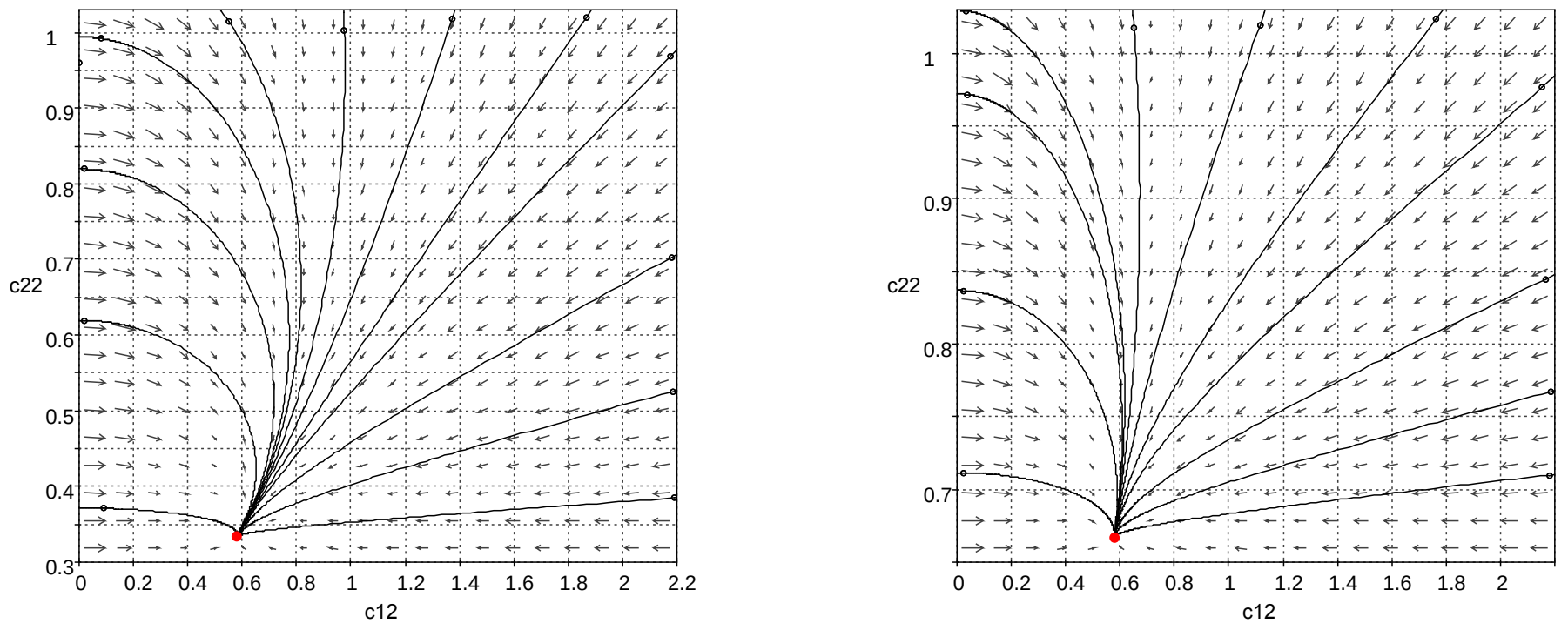

Figure 6: Phase plane plot of fast dynamics at $\beta = 0.5$ and $\beta = 2$.

## 3. The full Rolie-Poly model with $\delta = 0$

We now turn to the discussion of the full Rolie-Poly model. In general, this model seems too complicated to allow for analytical solutions, and we therefore focus on the case $\delta = 0$ in this section. The nonstretching version discussed above was based on taking the limit $\tau_R \to 0$, leaving only $\tau_d$ as a timescale. We introduced a retardation time in order to create a second timescale. For the full Rolie-Poly model, we already have two timescales, with ratio $\epsilon$. A retardation term would create a third timescale, and we do not consider it. One consequence of not having a retardation term is the occurence of step strains in the fast dynamics, which will be discussed below.

### *3.1. Steady shear flow*

For steady shear flow at a prescribed shear stress $C_{12} = \tau$, we obtain the following equations:

$$\begin{aligned}
2\kappa C_{12} - \epsilon(C_{11} - 1) - 2(1 - \sqrt{3/(C_{11} + 2C_{22})})(C_{11} + \beta(C_{11} - 1)) &= 0, \\
\kappa C_{22} - \epsilon C_{12} - 2(1 - \sqrt{3/(C_{11} + 2C_{22})})(1 + \beta)C_{12} &= 0, \\
-\epsilon(C_{22} - 1) - 2(1 - \sqrt{3/(C_{11} + 2C_{22})})(C_{22} + \beta(C_{22} - 1)) &= 0. \quad (36)
\end{aligned}$$

Figure 7 shows the steady flow curve for $\beta = 0.5$ and $\beta = 2$. In the following, we analyze the behavior of the steady flow curves in the limit of small $\epsilon$. There are two separate parts to the flow curve: an "unyielded" part where both the shear rate $\kappa$ and $\operatorname{tr}\mathbf{C} - 3$ are of order $\epsilon$, corresponding roughly to parts (a) and (c) of the figure, and a "yielded" part where $C_{22}$ is close to $\beta/(1+\beta)$, corresponding roughly to parts (b) and (d) of the figure.

We first consider the case where both the shear rate $\kappa$ and $\operatorname{tr}\mathbf{C} - 3$ are of order $\epsilon$. We therefore set $C_{11} = 3 - 2C_{22} + \epsilon d$, $\kappa = \epsilon\tilde{\kappa}$, and truncate the equations at order $\epsilon$. This leads to

$$\begin{aligned} 2\tilde{\kappa}C_{12} + 2C_{22} - 2 + \frac{d}{3}(-3 - 2\beta + 2C_{22}(1+\beta)) = 0, \\ \tilde{\kappa}C_{22} - C_{12} - \frac{d}{3}C_{12}(1+\beta) = 0, \\ 3 - 3C_{22} + d(\beta - C_{22}(1+\beta)) = 0. \end{aligned} \quad (37)$$

We can eliminate $d$ and $C_{12}$ from the last two equations. This leads to

$$\tilde{\kappa}^2 = \frac{3(1 - C_{22})}{2C_{22}((1+\beta)C_{22} - \beta)^2}, \quad (38)$$

and

$$C_{12}^2 = \frac{3}{2}C_{22}(1 - C_{22}). \quad (39)$$

This result is identical to that for the nonstretching Rolie-Poly model in the previous section. On the physically relevant branch, the shear rate increases from zero to infinity as $C_{22}$ decreases from 0 to $\beta/(1+\beta)$. The shear stress goes through a maximum when $C_{22} = 1/2$, if $\beta$ is less than 1. For $\beta > 1$, the shear stress is monotone. The limiting value at $C_{22} = \beta/(1+\beta)$ is $C_{12} = \sqrt{3\beta/2}/(1+\beta)$.

As $C_{22}$ approaches $\beta/(1+\beta)$, $\tilde{\kappa}$ tends to infinity, and the assumption that the shear rate is of order $\epsilon$ is no longer valid. For $C_{22}$ close to $\beta/(1+\beta)$, we therefore need a different analysis. We start with the last equation of (36); in this equation, we set $C_{22} = \beta/(1+\beta) + d$, and solve for $\operatorname{tr}\mathbf{C}$. The result is

$$\operatorname{tr}\mathbf{C} = \frac{12(1+\beta)^4 d^2}{(2d(1+\beta)^2 - \epsilon + (1+\beta)d\epsilon)^2}. \quad (40)$$

This is substituted into the second equation of (36). We can solve this equation for $\kappa$;

$$\kappa = \frac{C_{12}\epsilon}{d(\beta + d + \beta d)}. \quad (41)$$

We use the above expressions for $\kappa$ and $\operatorname{tr}\mathbf{C}$ and insert them into the first equation in (36). For small $\epsilon$ and $d$, the leading order balance becomes

$$2C_{12}^2(1+\beta)^2(-2(1+\beta)^2d+\epsilon)^2-3\beta(4(1+\beta)^4d^2+4\beta(1+\beta)^2\epsilon d-\beta\epsilon^2)=0. \quad (42)$$

This leads to the following conclusions:

1. A solution exists for $d > \epsilon/(2(1+\beta)^2)$. As $d$ tends to $\epsilon/(2(1+\beta)^2)$, $C_{12}$ approaches infinity. Moreover, we see from (41) that the viscosity $C_{12}/\kappa$ approaches the limit $\beta/(2(1+\beta)^2)$. In the opposite limit, as $d/\epsilon$ approaches infinity, $C_{12}$ approaches $\sqrt{3\beta/2}/(1+\beta)$.
2. A more careful analysis, which will be omitted here, shows that for $\beta < 1$, $C_{12}$ assumes a minimum when $d$ is of order $\sqrt{\epsilon}$, and $d^2$ is approximately $\epsilon\beta/((1-\beta)(1+\beta)^2)$.

Figure 8 shows the steady flow curves corresponding to Figure 7, computed using the asymptotic analysis discussed above. The agreement between the two figures is evident.

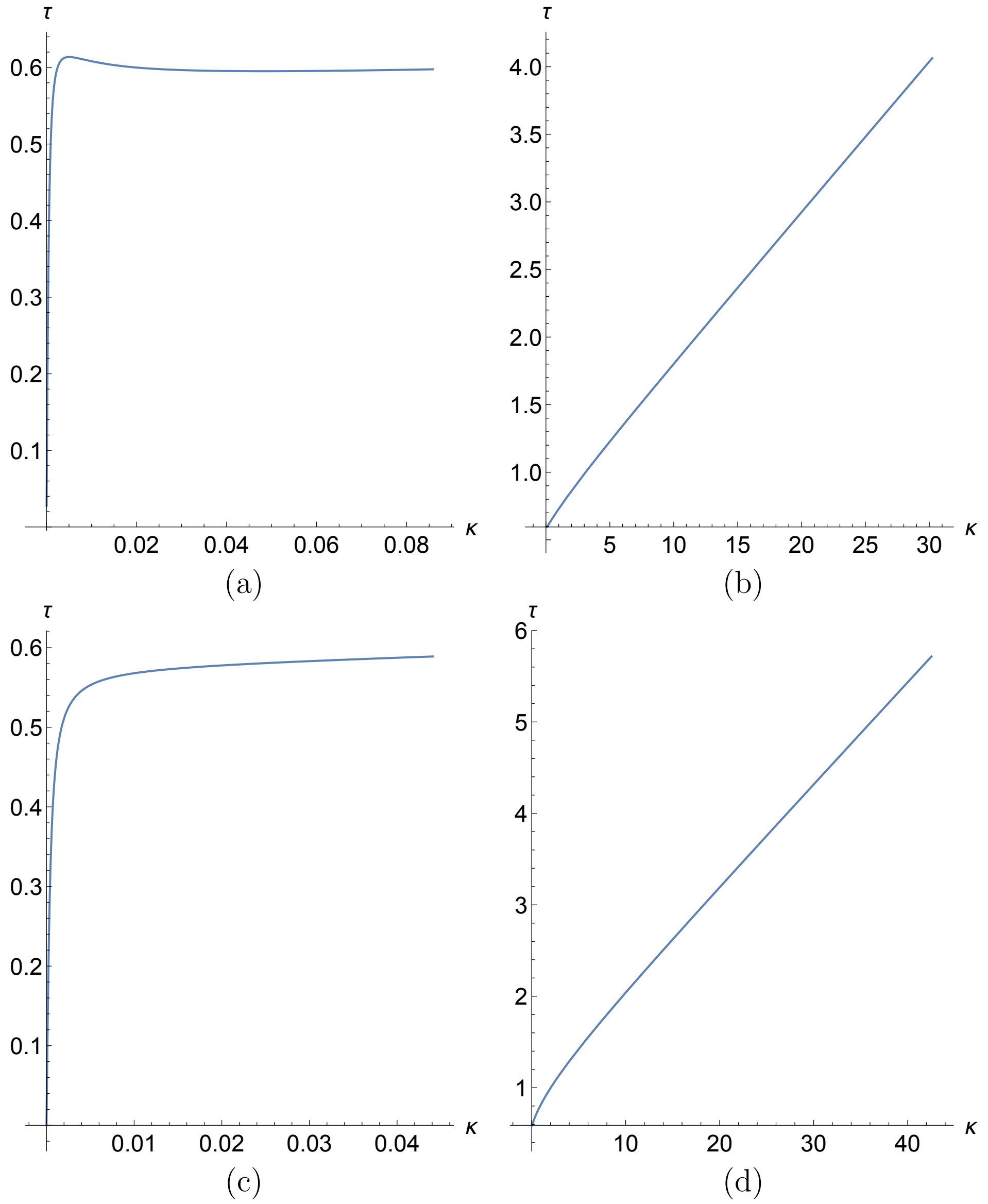

Figure 7: Steady flow curve for the full Rolie-Poly model, $\epsilon = 0.001$, $\delta = 0$. (a) $\beta = 0.5$, small shear rate, (b) $\beta = 0.5$, large shear rate, (c) $\beta = 2$, small shear rate, (b) $\beta = 2$, large shear rate.

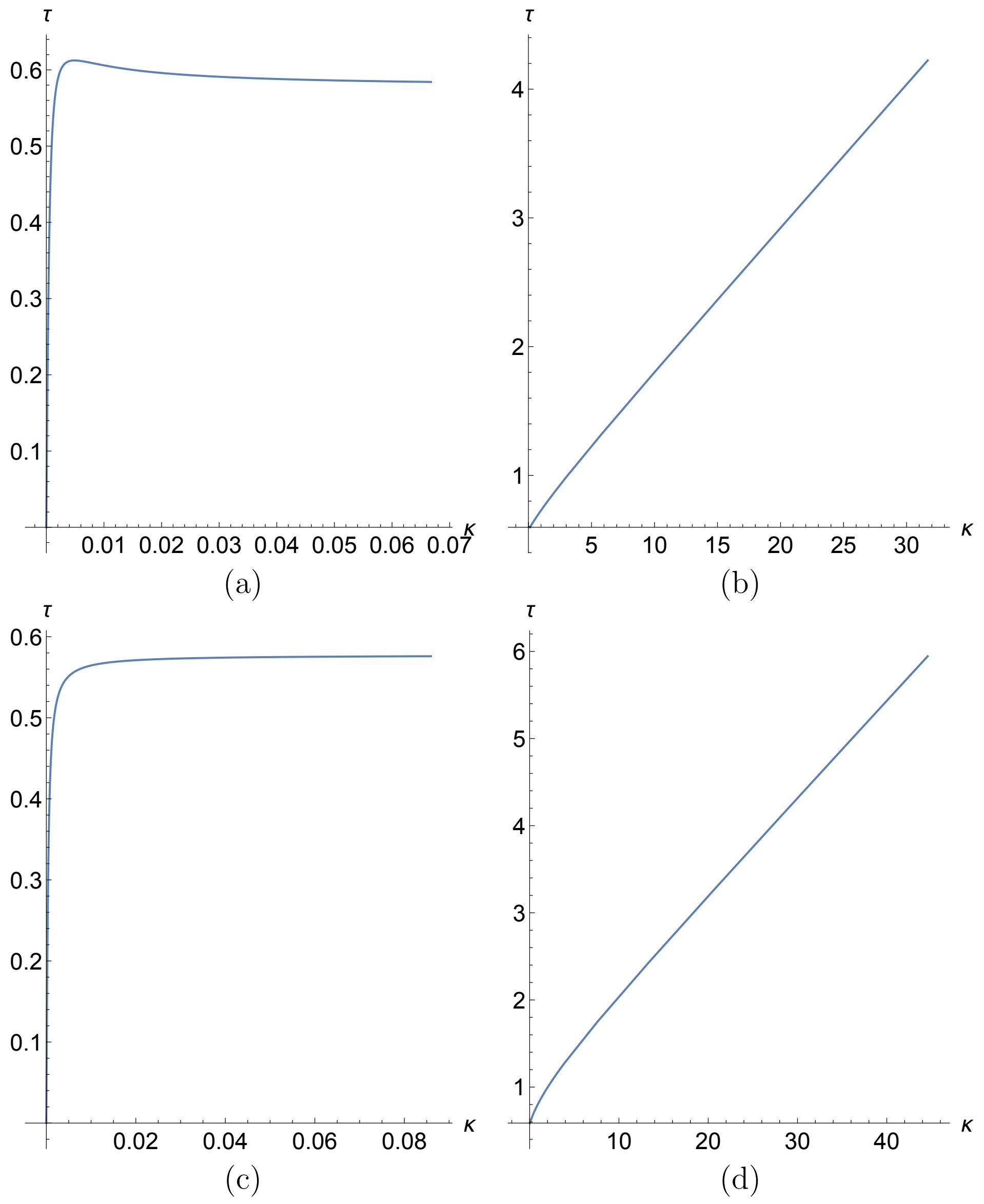

Figure 8: Steady flow curve for the full Rolie-Poly model, $\epsilon = 0.001$, $\delta = 0$, based on the asymptotic analysis of this section: (a) $\beta = 0.5$, small shear rate, (b) $\beta = 0.5$, large shear rate, (c) $\beta = 2$, small shear rate, (b) $\beta = 2$, large shear rate.

*3.2. Fast dynamics*

If we set $\epsilon = 0$, we obtain the following system of equations:

$$\begin{aligned}\dot{C}_{11} &= 2\kappa C_{12} - 2(1-\sqrt{3/\operatorname{tr}\mathbf{C}})(C_{11}+\beta(C_{11}-1)),\\ \dot{C}_{12} &= \kappa C_{22} - 2(1-\sqrt{3/\operatorname{tr}\mathbf{C}})(1+\beta)C_{12},\\ \dot{C}_{22} &= -2(1-\sqrt{3/\operatorname{tr}\mathbf{C}})(C_{22}+\beta(C_{22}-1)).\end{aligned} \quad (43)$$

We are interested in the startup of shear flow, where the shear stress $C_{12}$ jumps from 0 to $\tau$. That is, $C_{12}$ is a Heaviside function, and $\dot{C}_{12}$ is a delta function. The jump in stress induces a jump in strain, i.e. the shear rate $\kappa$ also becomes a delta function. We deduce from (43) that there is no jump in $C_{22}$. Moreover, we can solve the second equation for $\kappa$, and in the first equation we obtain

$$\dot{C}_{11} = \frac{2C_{12}\dot{C}_{12}}{C_{22}} + ... = \frac{1}{C_{22}}\frac{d}{dt}(C_{12}^2) + ..., \quad (44)$$

where the dots indicate terms that do not include a delta function. By integrating this, we conclude that the jump in $C_{11}$ is $1/C_{22}$ times the jump in $C_{12}^2$. If we start from rest and suddenly impose a shear stress $\tau$, we therefore have the initial condition that $C_{11} = 1 + \tau^2$, $C_{12} = \tau$, and $C_{22} = 1$.

We now consider the continuous evolution from this initial condition, assuming that $C_{12} = \tau$ is held fixed. It follows that

$$\kappa = 2(1-\sqrt{3/\operatorname{tr}\mathbf{C}})(1+\beta)\tau/C_{22}. \quad (45)$$

We use this in the first equation of (43). Moreover, we introduce a new time variable $u$ by setting $du/dt = 1 - \sqrt{3/\operatorname{tr}\mathbf{C}}$. Finally, we set $s = \operatorname{tr}\mathbf{C}$. In this fashion, we obtain the following equations:

$$\begin{aligned}\frac{dC_{22}}{du} &= -2(C_{22}+\beta(C_{22}-1)),\\ \frac{ds}{du} &= \frac{4}{C_{22}}(1+\beta)\tau^2 + 2(3\beta-(1+\beta)s).\end{aligned} \quad (46)$$

The first equation yields the solution

$$C_{22} = \frac{\beta}{1+\beta} + c_1 e^{-2(1+\beta)u}, \quad (47)$$

and the initial condition $C_{22} = 1$ yields $c_1 = 1/(1+\beta)$. We can now use this solution in the second equation. With the initial condition $s = 3 + \tau^2$, we find

$$s = \frac{3\beta}{1+\beta} + \frac{2\tau^2(1+\beta)}{\beta} + \frac{3}{1+\beta}e^{-2(1+\beta)u} - \frac{\tau^2(2+\beta)}{\beta}e^{-2(1+\beta)u} + \frac{2\tau^2(1+\beta)}{\beta^2}e^{-2(1+\beta)u}(\ln(1+\beta) - \ln(1+\beta e^{2(1+\beta)u})). \tag{48}$$

We now have two alternatives:

1. $s$ reaches the value 3 at a finite value of $u$. In that case, $\operatorname{tr}\mathbf{C} - 3$ and $\kappa$ become zero at this point. We have a transition to slow dynamics, discussed in the next subsection.
2. If $s$ does not reach the value 3, we have $C_{22} \to \beta/(1+\beta)$ and $s \to 3\beta/(1+\beta) + 2\tau^2(1+\beta)/\beta$ for $u \to \infty$. At this point, we have a transition to yielded dynamics.

Thus, everything hinges on whether $s$ as given by (48) does or does not reach the value 3 at a finite positive value of $u$. Unfortunately, it does not seem easy to determine this from (48) directly, and we pursue a different approach. We note that $s(0) > 3$. If the limit of $s$ for $u \to \infty$ is less than 3, then $s$ must definitely cross 3. This is the case if

$$\tau^2 < \frac{3\beta}{2(1+\beta)^2}. \tag{49}$$

If $\tau^2$ is bigger than this value, then both $s(0)$ and $s(\infty)$ are bigger than 3. The only way in which $s$ can reach 3 is if it goes through a minimum, and the value of the minimum is less than or equal to 3. We now look for the limiting case when the value of the minimum is exactly 3. If $s = 3$ and $ds/du = 0$, then we conclude from the second equation of (46) that

$$C_{22} = \frac{2}{3}(1+\beta)\tau^2. \tag{50}$$

We can now determine tha value of $u$ for which this is the case and use it in the expression for $s$. The result is a quite lengthy expression depending on $\beta$ and $\tau$, which we do not reproduce. Instead we show a contour plot showing the line where that expression is equal to 3.

The upper curve in Figure 9 is the line in the $(\beta, \tau)$ plane where the solution given by (48) reaches a minimum of 3; that is, below the curve $s$ reaches 3, above it does not. This curve, as shown below, marks the boundary between immediate and delayed yielding. The lower curve in the plot is the boundary between delayed yielding and unyielded behavior. As the next subsection shows, this is the same as for the nonstretching Rolie-Poly model.

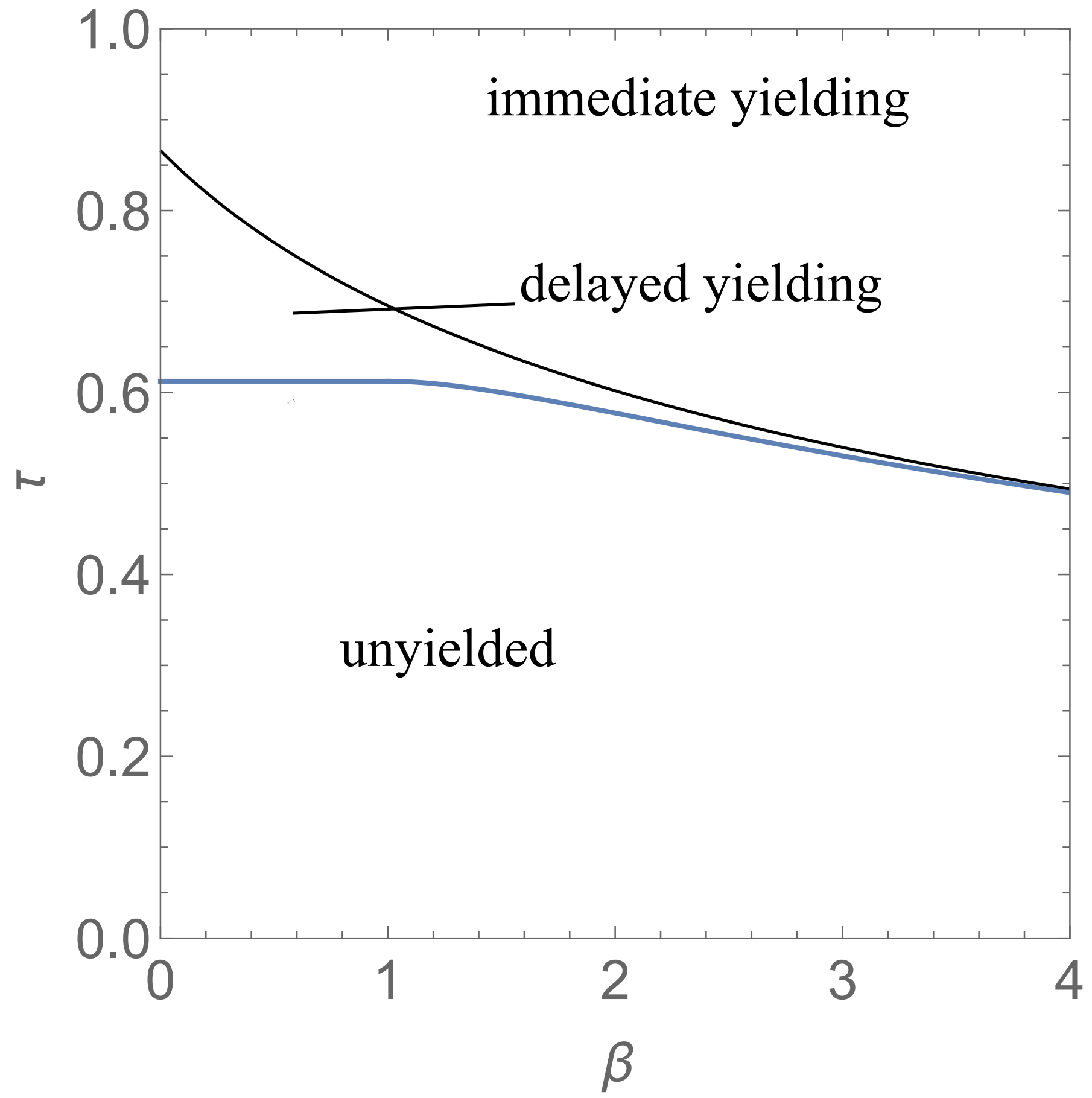

Figure 9: Regions of unyielded behavior, delayed yielding and immediate yielding for the full Rolie-Poly model, $\delta = 0$.

*3.3. Slow dynamics*

To analyze slow dynamics, we set $\operatorname{tr}\mathbf{C} = C_{11} + 2C_{22} = 3 + \epsilon d$, $\kappa = \epsilon\tilde{\kappa}$, and we rescale the time as $t = \tilde{t}/\epsilon$. At leading order, we obtain the equations

$$\begin{aligned} \frac{dC_{22}}{d\tilde{t}} &= 1 - C_{22} - \frac{d}{3}(C_{22} + \beta(C_{22} - 1)), \\ C_{22}\tilde{\kappa} &= \tau(1 + \frac{d}{3}(1 + \beta)), \\ d &= 2\tau\tilde{\kappa}. \end{aligned} \tag{51}$$

We find

$$\begin{aligned} d &= \frac{6\tau^2}{3C_{22} - 2(1+\beta)\tau^2}, \\ \tilde{\kappa} &= \frac{3\tau}{3C_{22} - 2(1+\beta)\tau^2}, \end{aligned} \tag{52}$$

and $C_{22}$ obeys the same equation as (22) derived above for the slow dynamics of the nonstretching Rolie-Poly model.

In contrast to the nonstretching case, there is no closed form expression for $C_{22}^i$, the initial value of $C_{22}$ where fast dynamics transitions to slow dynamics in startup of shear flow. However, some qualitative properties follow. First, the transition from fast to slow dynamics must occur at $s = 3$ and $ds/du < 0$ in (46). This implies that

$$C_{22}^i > C_{22}^s = 2(1+\beta)\tau^2/3, \tag{53}$$

where $C_{22}^s$ is the singular value defined in (24). In the nonstretching case, we showed that when there are equilibrium points, then $C_{22}^i$ is bigger than the larger of the two equilbrium values. We have no proof of this in the stretching case, but for our conclusions below, it suffices to show that $C_{22}^i > 1/2$, the average of the two equilibrium values. In conjunction with the above result that $C_{22}^i > C_{22}^s$, and the obvious fact that $C_{22}^i > \beta/(1+\beta)$, this guarantees that as long as there is a stable equilibrium value for $C_{22}$ in the range $1 > C_{22} > \beta/(1+\beta)$, the solution on the slow manifold will converge to that equilibrium value. We have $C_{22}^i > \beta/(1+\beta) > 1/2$ if $\beta > 1$. The remaining case is $\beta \le 1$ and $\tau^2 \le 3/8$. We consider the solution (47), determine $u$ such that $C_{22} = 1/2$, and compute the corresponding $s$. The result is

$$s = \frac{3}{2} + \tau^2 \frac{\beta(2 + 5\beta + \beta^2) + 2(1-\beta^2)\ln(1-\beta)}{2\beta^2}. \tag{54}$$

A contour plot (which we omit) shows that this value is less than or equal to 3 in the range $0 \le \beta \le 1$, $0 \le \tau^2 \le 3/8$, with equality only when $\beta = 1$, $\tau^2 = 3/8$. Therefore, the value $s = 3$ is reached before $C_{22}$ gets to $1/2$.

We can therefore obtain results for startup of steady shear flow which are analogous to the nonstretching case. Specifically, the behavior is as follows:

1. If either $\beta < 1$ and $\tau < \sqrt{3/8}$ or $\beta \ge 1$ and $\tau < \sqrt{3\beta/2}/(1+\beta)$, an unyielded equilibrium is reached via slow dynamics.
2. If $\beta < 1$, $\tau > \sqrt{3/8}$ or $\beta \ge 1$, $\tau > \sqrt{3\beta/2}/(1+\beta)$, but $\tau$ is still less than the value given by the upper curve in Figure 9, delayed yielding occurs.
3. If $\tau$ is larger than the value given by the upper curve in Figure 9, immediate yielding occurs.

*3.4. Yielded dynamics*

We now study the dynamics when $C_{22}$ is close to $\beta/(1+\beta)$. We set $C_{22} = \beta/(1+\beta) + \epsilon q$. At leading order in $\epsilon$, we obtain the reduced equation

$$\frac{\beta\kappa}{1+\beta} = 2(1 - \sqrt{3/s})(1+\beta)\tau. \tag{55}$$

Using this in the equation for the evolution of $s$, we find

$$\frac{ds}{dt} = -2(1-\sqrt{3/s})(s+\beta(s-3)) + 4(1-\sqrt{3/s})\frac{(1+\beta)^2\tau^2}{\beta}. \tag{56}$$

Finally, we find

$$\frac{dq}{dt} = \frac{1}{1+\beta} - 2(1-\sqrt{\frac{3}{s}})(1+\beta)q. \tag{57}$$

The equation (56) has the equilibrium points $s = 3$ and

$$s_0 = \frac{3\beta^2 + 2\tau^2(1+\beta)^2}{\beta(1+\beta)}. \tag{58}$$

If $\tau^2 > 3\beta/(2(1+\beta)^2)$, $s_0 > 3$. Moreover, $s_0$ is stable, and $s = 3$ is unstable. If $\tau^2 < 3\beta/(2(1+\beta)^2)$, then $s_0 < 3$, and the stability becomes reversed. If we start from a yielded steady shear flow and then lower the value of $\tau$, the flow will remain yielded if $\tau^2 > 3\beta/(2(1+\beta)^2)$, but it will unyield and transition to slow dynamics if $\tau^2$ is less than this value.

*3.5. Thixotropy*

As for the unstretched version of the model above, we can observe thixotropic behavior. Suppose we start from a yielded shear flow and then reduce $\tau$ to zero. On the slow manifold, $s$ is close to 3, $\kappa$ is zero, but $C_{22}$ grows from $\beta/(1+\beta)$ back to 1 only slowly, on a timescale of $1/\epsilon$. If we now reimpose a shear stress, then the initial jump of $s$ will be equal to $\tau^2/C_{22}$, which is larger than it would be if we started from equilibrium. Also, the shear rate is given by

$$\kappa = \frac{2(1-\sqrt{3/s})(1+\beta)+\epsilon}{C_{22}}\tau, \tag{59}$$

so the apparent viscosity $\tau/\kappa$ decreases with increasing $s$ and decreasing $C_{22}$. On the slow manifold, according to (52), the apparent viscosity is

$$\frac{3C_{22}-2(1+\beta)\tau^2}{3\epsilon}. \tag{60}$$

*3.6. The case $\beta = 0$*

So far, we have always assumed that $\beta > 0$. We conclude this section with a discussion of the special case $\beta = 0$. The nonstretching case, with the added Newtonian term, is equivalent to the PEC model discussed in [6]; we note that the tensor denoted by $\mathbf{C}$ there is not the same as in this paper: our $\mathbf{C}$ would correspond to $\mathbf{C}/s$ in the notation of [6]. We now give a discussion of the stretching case.

The difference to the case of nonzero $\beta$ is that the value of $C_{22}$ in the yielded regime becomes of order $\epsilon$. To focus on this regime, we investigate what happens when $\epsilon$ and $C_{22}$ are small, and $s-3$ is not of order $\epsilon$, but larger. For steady flow, we obtain the following simplified set of equations:

$$\begin{aligned}
2\kappa C_{12} - 2(1-\sqrt{\frac{3}{s}})s &= 0,\\
\kappa C_{22} - 2(1-\sqrt{\frac{3}{s}})C_{12} &= 0,\\
\epsilon - 2(1-\sqrt{\frac{3}{s}})C_{22} &= 0.
\end{aligned} \tag{61}$$

We can solve for $C_{12}$, $C_{22}$ and $\kappa$ in terms of $s$. The result is

$$C_{22} = \frac{\epsilon}{2(1-\sqrt{3/s})},$$

$$
\begin{aligned}
C_{12}^2 &= \frac{\epsilon s}{4(1-\sqrt{3/s})}, \\
\kappa &= \frac{4C_{12}(1-\sqrt{3/s})^2}{\epsilon}.
\end{aligned} \tag{62}
$$

Thus the limiting viscosity at high shear rate ($s \to \infty$) is $\epsilon/4$. A minimum of shear stress is reached when $s = 27/4$; the corresponding shear rate and shear stress are $\kappa = \sqrt{1/\epsilon}$ and $C_{12} = \frac{9}{4}\sqrt{\epsilon}$.

To study yielded dynamics with an imposed shear stress $\tau$ of order 1, we set $s = \tilde{s}/\epsilon$, $C_{22} = c_{22}\epsilon$, and $\kappa = \tilde{\kappa}/\epsilon$. At leading order, we find the equations

$$
\begin{aligned}
\dot{\tilde{s}} &= 2\tilde{\kappa}\tau - 2\tilde{s}, \\
\dot{c}_{22} &= 1 - 2c_{22}, \\
\kappa c_{22} &= 2\tau.
\end{aligned} \tag{63}
$$

These equations have a stable fixed point at $c_{22} = 1/2$, $\tilde{s} = 2\tau^2$. Unyielding will not occur unless the shear stress is lowered to zero (more precisely, to $O(\sqrt{\epsilon})$).

Now we consider what happens during unyielding. We note that $C_{22}$ obeys the equation

$$
\dot{C}_{22} = \epsilon(1 - C_{22}) - 2(1 - \sqrt{\frac{3}{s}})C_{22}. \tag{64}
$$

If we start with an initial value of order $\epsilon$, then $C_{22}$ will not grow beyond order $\epsilon$ until $s - 3$ becomes small, i.e. slow dynamics has been reached. During the slow phase of unyielding, $C_{22}$ therefore grows from order $\epsilon$ back to its equilibrium value of 1, i.e. it changes by a factor of $1/\epsilon$ rather than a factor of order 1 as it did for nonzero $\beta$. Hence the model with $\beta = 0$ is much more thixotropic than the model with nonzero $\beta$.

## 4. Numerical results for nonzero $\delta$

### *4.1. Steady shear flow*

For steady shear flow, we have the equations

$$
\begin{aligned}
-\epsilon(C_{22} - 1) - 2(1 - \sqrt{s/3})(C_{22} + \beta(\frac{s}{3})^\delta(C_{22} - 1)) &- 0, \\
\kappa C_{22} - \epsilon C_{12} - 2(1 - \sqrt{s/3})(1 + \beta(\frac{s}{3})^\delta)C_{12} &= 0, \\
2\kappa C_{12} - \epsilon(s - 3) - 2(1 - \sqrt{s/3})(s + \beta(\frac{s}{3})^\delta(s - 3)) &= 0.
\end{aligned} \tag{65}
$$

We can solve the first equation of (65) for $C_{22}$, the result is

$$C_{22} = \frac{\epsilon + 2\beta(s/3)^\delta(1-\sqrt{3/s})}{\epsilon + 2(1-\sqrt{3/s})(1+\beta(s/3)^\delta)}. \tag{66}$$

We can then solve the second equation of (65) for $C_{12}$, use the result in the first equation, and then solve for $\kappa$. This will yield both $C_{12}$ and $\kappa$ as functions of $s$. This procedure was used to produce the flow curves which are shown below. We do not reproduce the formulae, which are quite complicated, but we discuss some qualitative aspects.

If $s$ is close to 3, then there is little difference between $(s/3)^\delta$ and 1. Thus, the slow dynamics of the previous section is independent of $\delta$. We now discuss the steady flow behavior in the opposite limit, when $s/3$ is large. If $\delta$ is positive, we obtain $C_{22} \sim 1$ in this limit, and the remaining equations simplify to

$$\begin{aligned} \kappa &\sim 2\beta(\frac{s}{3})^\delta C_{12}, \\ 2\kappa C_{12} &\sim 2\beta s(\frac{s}{3})^\delta. \end{aligned} \tag{67}$$

This leads to

$$\begin{aligned} C_{12} &\sim \sqrt{\frac{s}{2}}, \\ \kappa &\sim \frac{\sqrt{2}\beta}{3^\delta} s^{1/2+\delta}. \end{aligned} \tag{68}$$

Hence $C_{12}$ is proportional to $\kappa^{1/(1+2\delta)}$, i.e. we have shear thinning behavior.

For $\delta = 0$, we noted in the previous section that the viscosity approaches the constant value $\beta/(2(1+\beta)^2)$ at large shear rates.

If $\delta$ is negative, then (66) leads to different behaviors depending on whether $(s/3)^\delta$ is large or small relative to $\epsilon$. If $s$ is large, but still small relative to $\epsilon^{1/\delta}$, then

$$C_{22} \sim \beta(s/3)^\delta. \tag{69}$$

We then find

$$\begin{aligned} \kappa\beta(s/3)^\delta &\sim 2C_{12}, \\ 2\kappa C_{12} &\sim 2s. \end{aligned} \tag{70}$$

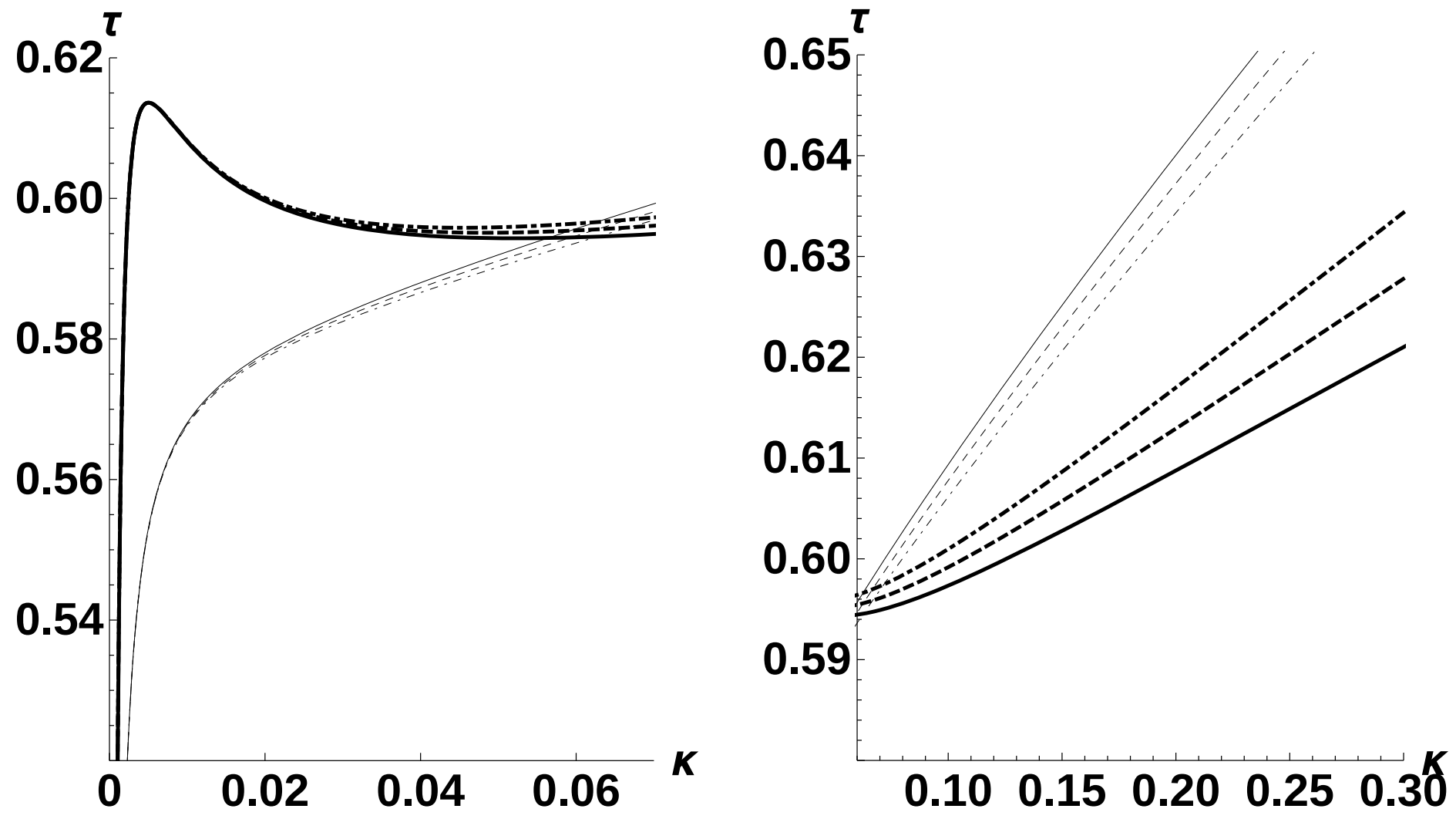

Figure 10: Steady flow curves for small shear rate and nonzero $\delta$. $\delta = -0.5$, $\beta = 0.5$ (**-**) and 2.0 (—); $\delta = 0$, $\beta = 0.5$ (**- - -**) and 2.0 ( - - -); $\delta = 0.5$, $\beta = 0.5$ (**- · -**) and 2.0 (- · -).

The result is a shear thinning behavior where $\kappa$ is proportional to $s^{(1-\delta)/2}$, and $C_{12}$ is proportional to $s^{(1+\delta)/2}$. On the other hand, if $s$ is large relative to $\epsilon^{1/\delta}$, then $C_{22}$ is close to $\epsilon/2$. The leading order balance in the remaining equations then becomes

$$\begin{aligned} \frac{\epsilon}{2}\kappa &\sim 2C_{12}, \\ 2\kappa C_{12} &\sim 2s. \end{aligned} \tag{71}$$

That is, we have a constant viscosity equal to $\epsilon/4$.

For the numerical simulations, we compare the values $\delta = -0.5$, $\delta = 0$ and $\delta = 0.5$. We focus on two values of $\beta$, one of which is chosen less than 1 and the other greater than 1; specifically we choose $\beta = 0.5$ and $\beta = 2$. Throughout, we set $\epsilon = 0.001$; this value was also chosen in [1] on the basis of a fit to experimental data for polybutadiene reported in [7]. Figure 10 shows the behavior at low shear rates. As expected, there is little dependence on $\delta$. For $\beta = 0.5$, the flow curve is nonmonotone, for $\beta = 2$ it is monotone, as predicted by the analysis in the preceding section.

For the behavior at large shear rate, we focus on $\beta = 0.5$; the results for $\beta = 2$ are similar. Figure 11 shows results for $\delta = 0$ and $\delta = 0.5$. For $\delta = 0$, we see linear behavior (constant viscosity), while for $\delta = 0.5$ we have shear

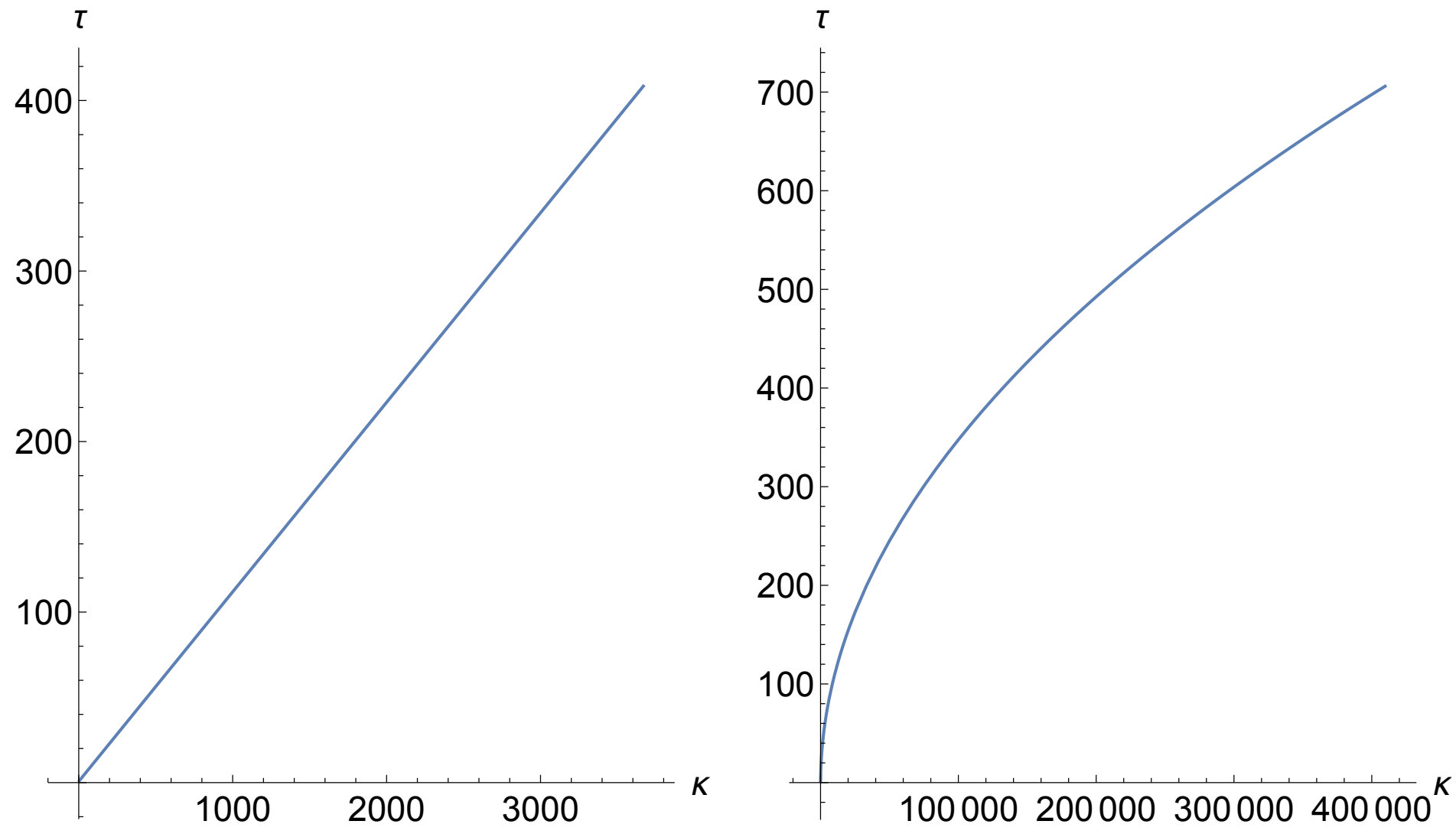

(a) (b)

Figure 11: Steady flow curves for large shear rate, $\beta = 0.5$. (a) For $\delta = 0$, the viscosity is constant. (b) For $\delta = 0.5$, the fluid is shear thinning.

thinning as predicted above.

For $\delta = -0.5$, the behavior is shown in Figure 12. We see shear thinning over a wide range, but eventually the slope becomes constant in accordance with the analysis above.

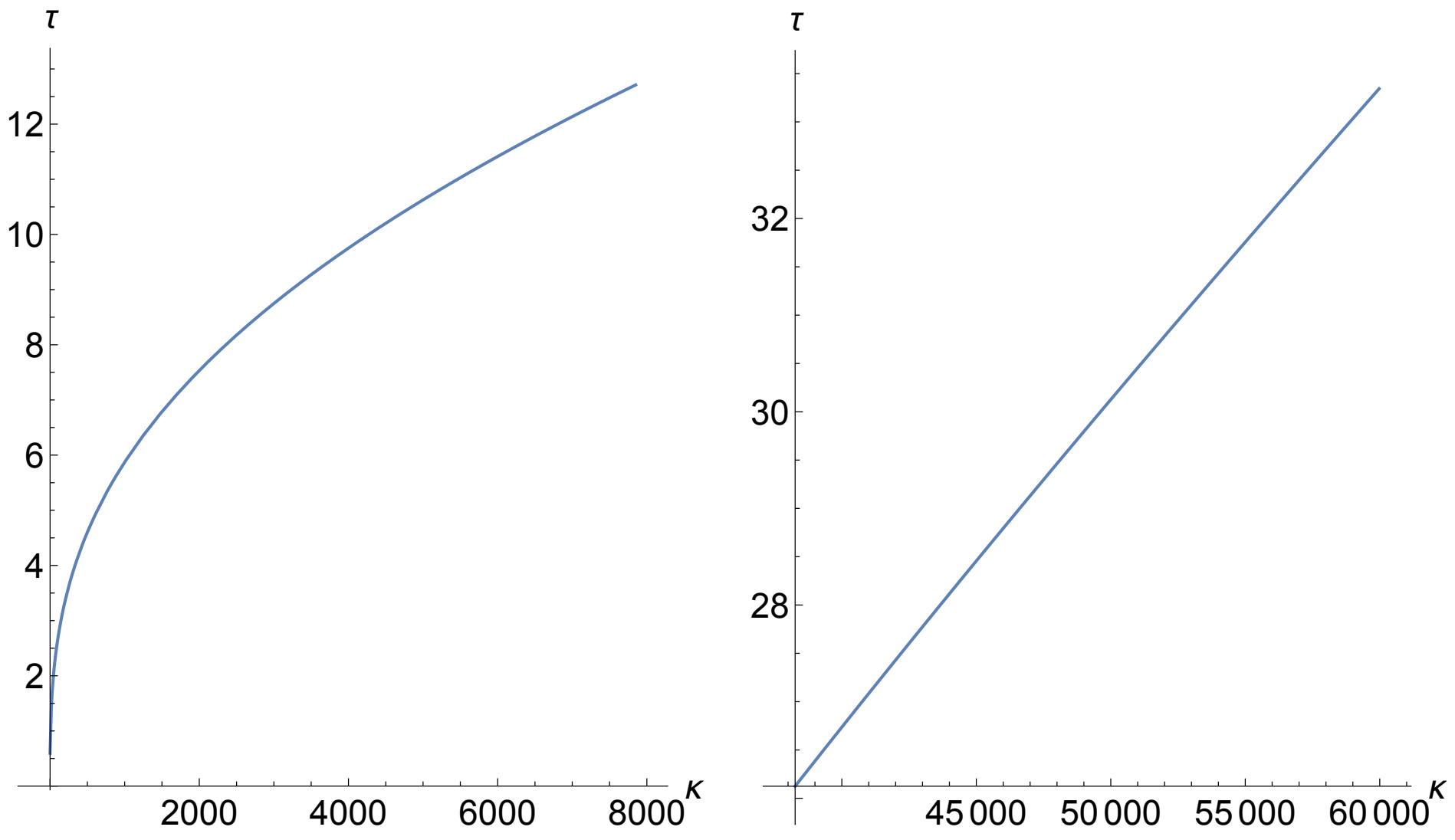

(a) (b)

Figure 12: Steady flow curves for large shear rate, $\beta = 0.5$, and $\delta = -0.5$. (a) The fluid is shear thinning over a wide range, but (b) eventually the viscosity becomes constant.

### *4.2. Startup of shear flow*

We now consider the startup of shear flow, i.e. the fluid is initially at rest, and for $t \geq 0$ a shear stress $\tau$ is imposed. Figure 13 shows the evolution of the shear rate $\kappa$ for $\beta = 0.5$, $\delta = -0.5$, and $\tau$ in the range $0.6 \leq \tau \leq 1$. The value 0.6 is less than $\sqrt{3/8}$ (the value of the steady shear stress maximum), and the flow settles to an unyielded steady state. As $\tau$ is increased, delayed yielding occurs. The delay shortens with increasing $\tau$, and eventually yielding becomes immediate.

Figure 14 shows similar behavior for $\delta = -0.5$ and $\beta = 2$. In this case, the maximum value of $\tau$ for unyielded steady flow is $\sqrt{3}/3 = 0.577$. Again we find delayed yielding in a window above this value.

Figures 15 and 16 show similar trends for $\delta = 0.5$.

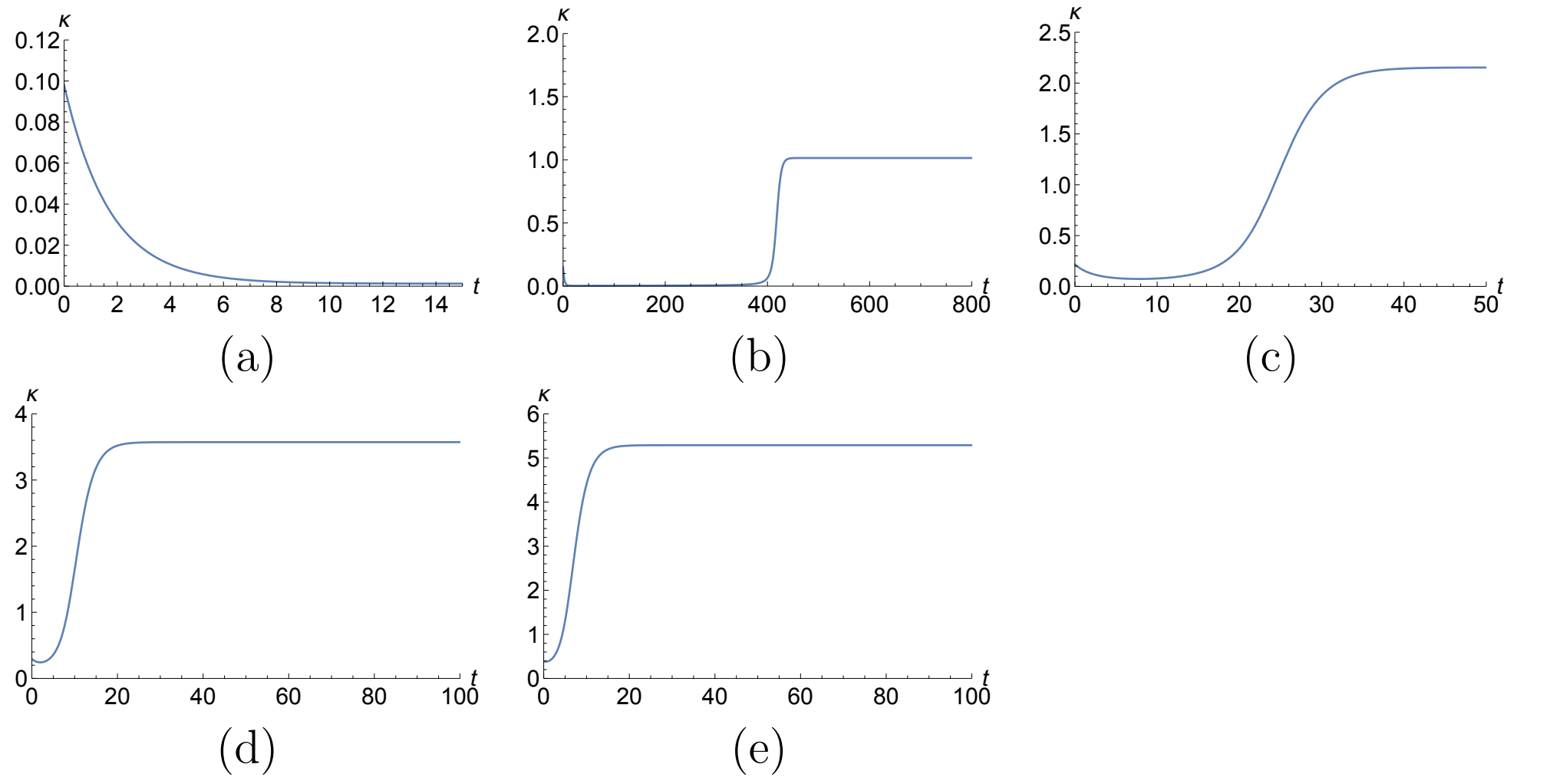

Figure 13: Evolution of shear rate for startup of shear flow, $\beta = 0.5$, $\delta = -0.5$. (a) $\tau = 0.6$, (b) $\tau = 0.7$, (c) $\tau = 0.8$, (d) $\tau = 0.9$, (e) $\tau = 1.0$.

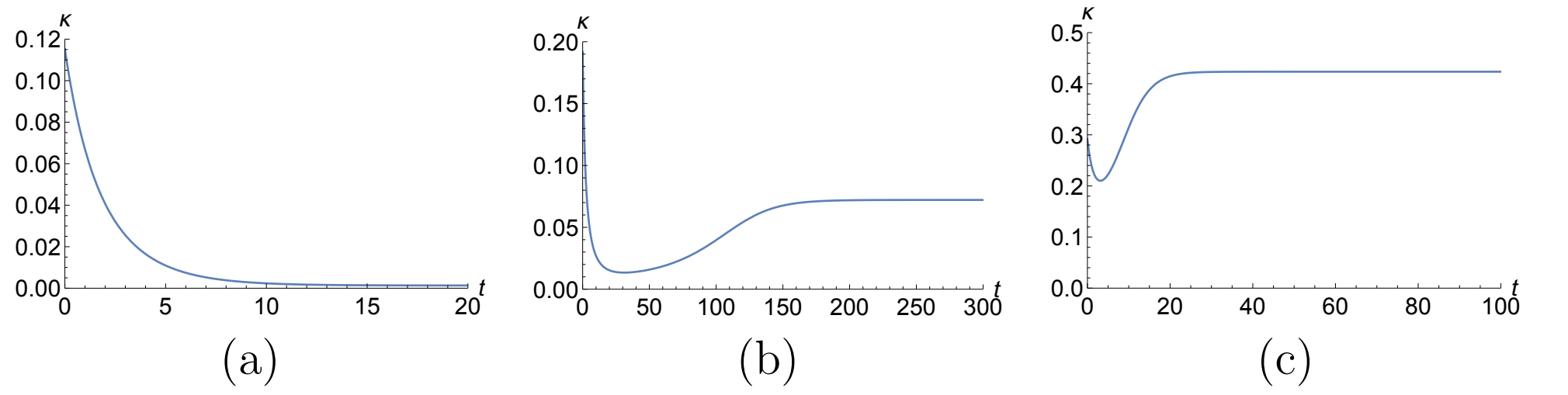

Figure 14: Evolution of shear rate for startup of shear flow, $\beta = 2$, $\delta = -0.5$. (a) $\tau = 0.5$, (b) $\tau = 0.6$, (c) $\tau = 0.7$.

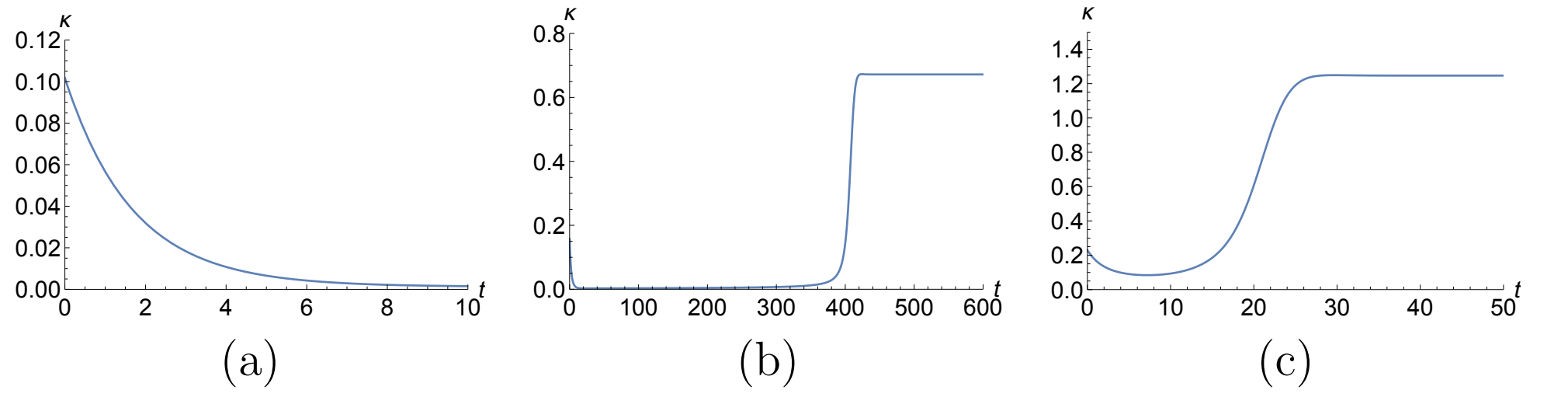

Figure 15: Evolution of shear rate for startup of shear flow, $\beta = 0.5$, $\delta = 0.5$. (a) $\tau = 0.6$, (b) $\tau = 0.7$, (c) $\tau = 0.8$.

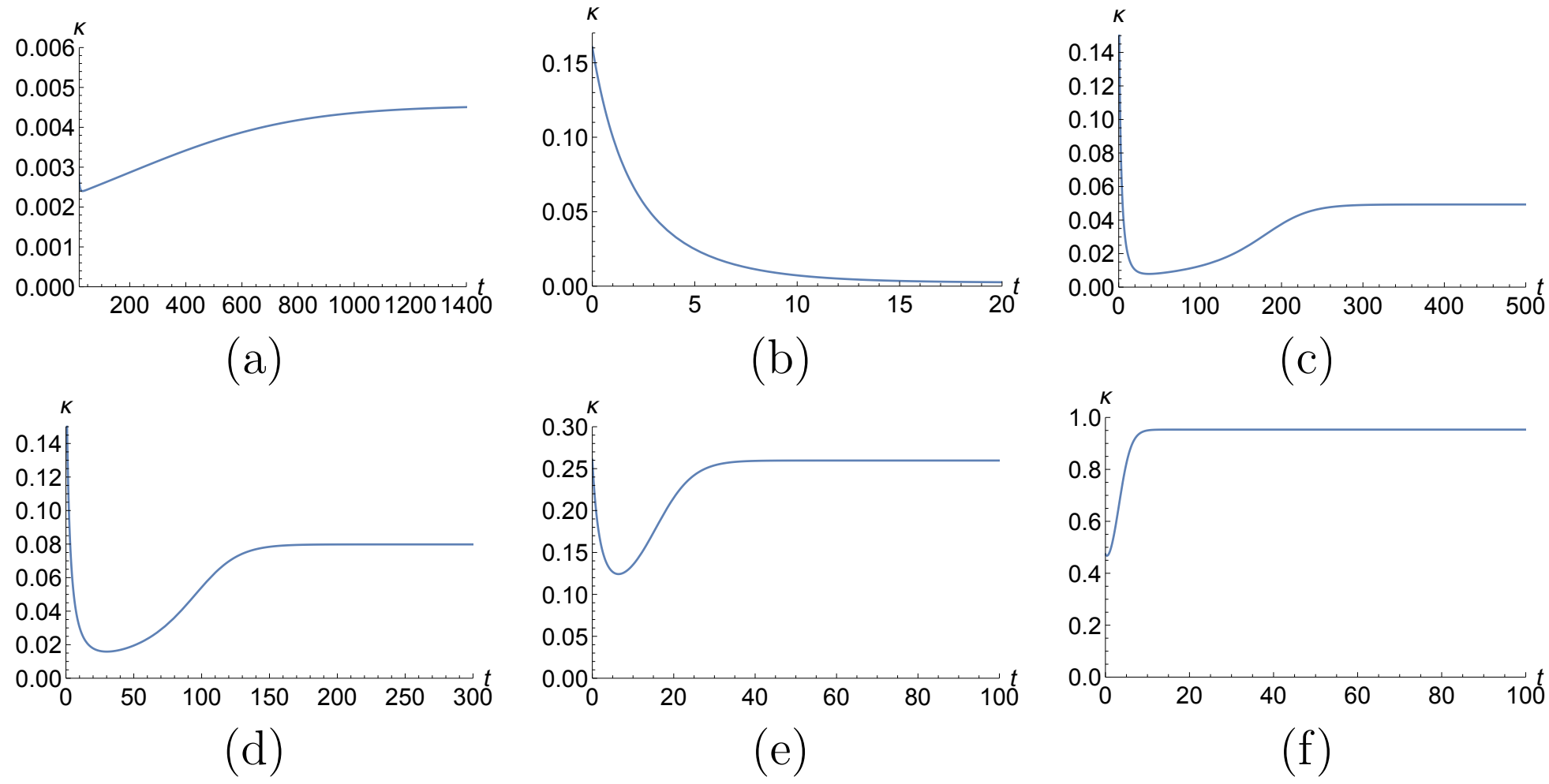

Figure 16: Evolution of shear rate for startup of shear flow, $\beta = 2$, $\delta = 0.5$. (a) $\tau = 0.55$, (b) $\tau = 0.55$, short times, (c) $\tau = 0.59$, (d) $\tau = 0.6$, (e) $\tau = 0.65$, (f) $\tau = 0.8$.

### 4.3. Unyielding and thixotropy

We show one instance of unyielding from an established steady flow. For the parameters, we choose $\beta = 0.5$, $\delta = -0.5$ and $\epsilon = 0.001$. We start with steady flow at $\tau = 1$, corresponding to $s = 9.87979$ in the parameter representation discussed at the beginning of Section 4.1. We then instantaneously reduce the imposed shear stress to zero. The initial conditions after the instantaneous step strain are $C_{11} = 4.83155$ and $C_{22} = 0.21669$. Figure 17 shows the time evolution of $s = C_{11} + 2C_{22}$ and $C_{22}$. As expected from the analysis above for $\delta = 0$ (see Section 3.5), $s$ relaxes quickly to 3, while $C_{22}$ reaches its equilibrium value of 1 only on a timescale of order $1/\epsilon$.

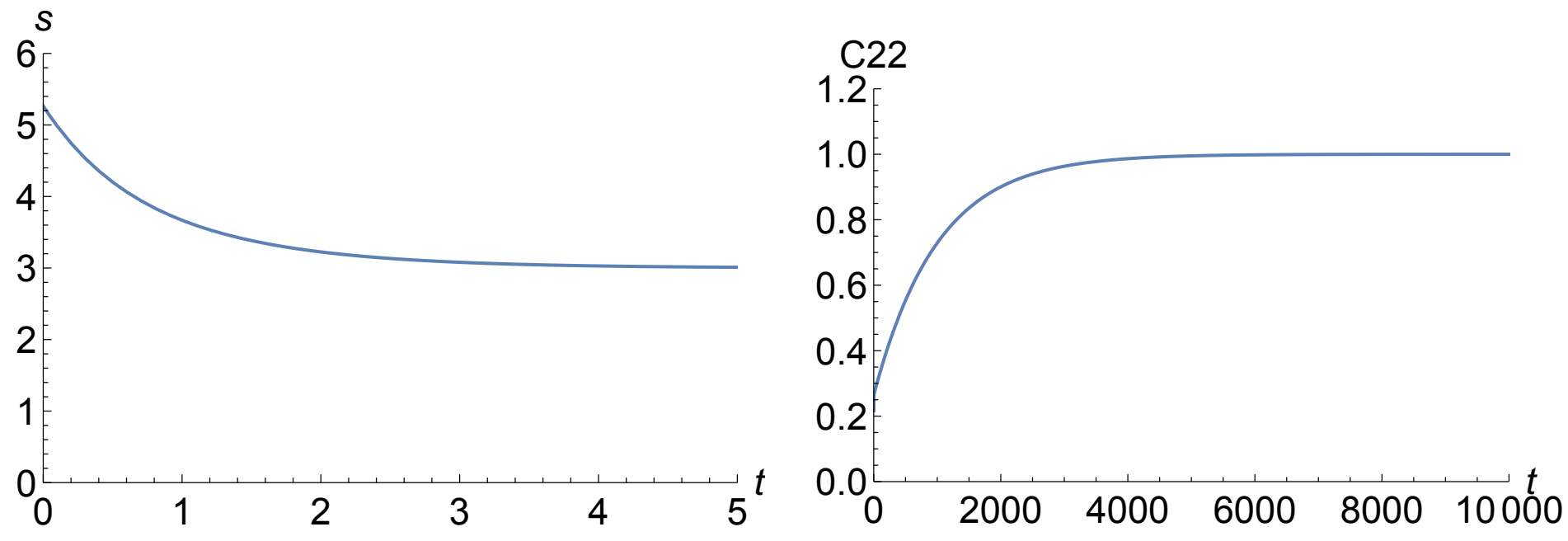

(a) (b)

Figure 17: The evolution of $s$, (a) and $C_{22}$, (b) after removal of the imposed shear stress. Unyielding from established steady shear flow, $\beta = 0.5$, $\delta = -0.5$. The initial shear flow is for $\tau = 1$.

## Conclusions

The Rolie-Poly model for entangled polymers features two distinct relaxation times, a Rouse time and a reptation time. In this paper, we have given an analysis, based on methods of singular perturbation theory, for the limit when the ratio of Rouse time to reptation time is small. We have first discussed the nonstretching limit, where the limit of zero Rouse time is taken. In this case, we introduced a second time scale by including a Newtonian viscosity, and thus a retardation time. We then analyzed the full Rolie-Poly model, in the special case where the exponent $\delta$ is zero. We identified regimes of fast, slow and yielded dynamics. Startup of shear flow leads to unyielded flow at low shear stress, immediate yielding at high shear stress, and a regime of delayed yielding at intermediate values. We establish specific criteria for these regimes in terms of the convective constraint release parameter $\beta$ and the imposed shear stress. In cessation of shear flow, the disparate time scales for relaxation to equilibrium lead to thixotropic behavior. A difference between the cases of zero and nonzero $\delta$ is that the high shear rate behavior for nonzero $\delta$ is shear-thinning. Shear-thinning behavior is found for both positive and negative $\delta$, the difference being that at very high shear rates, a limiting viscosity is reached when $\delta < 0$. Numerical results show that the transient behavior in startup and cessation of shear flow is similar to the more analytically tractable case of $\delta = 0$.

**Acknowledgment**

This research was supported by the National Science Foundation under Grants DMS-1311707 and DMS-1514576. The authors would like to thank the Department of Mathematics, University of British Columbia, for their hospitality during the time when this manuscript was written.